\documentclass[amsmath,amssymb,amsbsy,prb,twocolumn,superscriptaddress
]{revtex4-1}
\usepackage{graphicx}
\usepackage{braket}
\usepackage{bm}
\usepackage{color}
\usepackage{ulem}
\usepackage[colorlinks=true, citecolor=magenta, linkcolor=blue ]{hyperref}

\begin{document}

\newcommand{\txtr}[1]{\textcolor{red}{#1}}
\newcommand{\txtrout}[1]{\textcolor{red}{\sout{#1}}}
\newcommand{\txtb}[1]{\textcolor{blue}{#1}}
\newcommand{\txtg}[1]{\textcolor{green}{#1}}
\newcommand{\txtm}[1]{\textcolor{magenta}{#1}}

\newcommand{\tu}{t_1^{\rm U}}
\newcommand{\td}{t_1^{\rm D}}

\title{Flat-band engineering in tight-binding models: Beyond the
    nearest-neighbor hopping}

\author{Tomonari Mizoguchi}
\affiliation{Department of Physics, University of Tsukuba, 1-1-1 Tennoudai, Tsukuba, Ibaraki 305-8571, Japan}
\email{mizoguchi@rhodia.ph.tsukuba.ac.jp}

\author{Masafumi Udagawa}
\affiliation{Department of Physics, Gakushuin University, 1-5-1 Mejiro, Toshima-ku, Tokyo 171-8588, Japan}

\begin{abstract}
In typical flat-band models, defined as nearest-neighbor tight-binding models, 
flat bands are usually pinned to
the special energies, such as top or bottom of dispersive bands, or band crossing points. 
In this paper, we propose a simple method to tune 
the energy of flat bands without losing the exact flatness of the bands. 
The main idea is to add farther-neighbor hoppings to the original nearest-neighbor models, in such a way that the transfer integrals depend only on the Manhattan distance. 
We apply this method to several lattice models including the two-dimensional kagome lattice and the three dimensional pyrochlore lattice,
as well as their breathing lattices and non-line graphs. 
The proposed method will be useful for engineering flat bands to generate desirable properties, 
such as enhancement of $T_c$ of superconductors and nontrivial topological orders.  

\end{abstract}
\maketitle

\section{Introduction}
Diversity of materials may be attributed to the diversity of band structures.
The variety of band structures associated with lattice structures and orbital characters is a source of 
rich phenomena in condensed-matter systems, such as 
spin and orbital magnetism~\cite{Stoner1938,Kanamori1963,Ogata2015}, superconductivity~\cite{Sigrist1991,Mackenzie2003,Stewart2011}, topological insulators~\cite{Hasan2010,Qi2011}, and topological Dirac and Weyl semimetals~\cite{Vafek2014,Armitage2018}. 

Among characteristic band structures, 
a completely dispersionless band, in entire Brillouin zone, is called a flat band. 
One of the remarkable consequences of this \lq \lq quench" of kinetic energy is the emergence of a ferromagnetic ground state 
when introducing the Hubbard interactions, and there has been a long history of study in this context~\cite{Lieb1989,Mielke1991,Mielke1991_2,Tasaki1992,Mielke1993,Kusakabe1994,Aoki1994,Tasaki1998,Katsura2010}. 
Topological physics in exact and nearly flat-band systems also attracts considerable interests~\cite{Katsura2010,Tang2011,Sun2011,Neupert2011,Sheng2011,Wang2011,Liu2012,Bergholtz2015,Peotta2015,Liang2017,Misumi2017}. 
To study such intriguing physics associated with
the flat-band systems,
a number of tight-binding Hamiltonians, which mostly consider the nearest-neighbor (NN) hoppings, have been proposed. 

Quite recently, the possibility of flat-band-assisted superconductivity has been revisited
in correlated electron systems, where the interband scattering between dispersive and flat bands plays an essential role~\cite{Imada2000,Kuroki2005,Kobayashi2016,Matsumoto2018}.
In particular, this mechanism is
thought of as one of the possible origins of 
enhancement of $T_c$ in a twisted bilayer graphene with so-called ``magic angles"~\cite{Cao2018,Cao2018_2,Volovik2018,Yuan2018,Koshino2018,Ochi2018,Zou2018}.
There, it has been pointed out that the preferable band structure for such mechanism is 
(i) the flat band is located slightly above or below the Fermi level, and (ii) the dispersive band has a large density of states (DOS) nearby the flat band. 
Therefore, for further development of this mechanism for the high-$T_c$ superconductivity,
it is desirable to have an engineering method not only to realize flat band
but also to tune its energy. 

In the present paper, we propose a simple guiding principle to tune the energy of flat bands. 
It may sound surprising, since a flat band is extremely fragile; an infinitesimal amount of perturbation is enough to destroy its flatness~\cite{Slot2017}.
Nevertheless, we will show that it is possible to systematically control its energy while keeping its exact flatness.

The main idea is to add farther-neighbor hoppings to the usual NN models with flat band(s), 
in such a way that the transfer integrals depend only on the {\it Manhattan distance.} 
After this modulation, the resulting Hamiltonian is expressed by the polynomial of the original 
NN Hamiltonian. 
As a result, the eigenfunctions remain exactly the same as the original ones and only the dispersion relations and the flat-band energy are modulated.
Our method, due to its simplicity, has two prominent advantages:
(i) the flat bands retain exact flatness after the modulation of the Hamiltonian,
and (ii) we only need a few parameters to control a flat-band energy. 

The rest of this paper is organized as follows. 
In Sec. \ref{sec:formulation}, we explain the basic mechanism of our method. 
Then, in Sec. \ref{sec:result1}, we apply this method to the line graphs in two- and three-dimensions, where
the existence of flat band(s) in the NN hopping models is guaranteed~\cite{Mielke1991}.
In Sec. \ref{sec:result2} we apply this method with slight modifications 
to the breathing lattices and a class of Lieb lattices, which have the site or bond inhomogeneity.
Section \ref{sec:circuit} is devoted to the application of our method to an artificial material, namely, an electric circuit.
Finally, our conclusion is presented in Sec. \ref{sec:conclusion}. 
Some analytical formulas for the dispersions relations are shown in the appendix. 

\section{Formulation \label{sec:formulation}}
\begin{figure}
\centering
\includegraphics[width=0.98\linewidth]{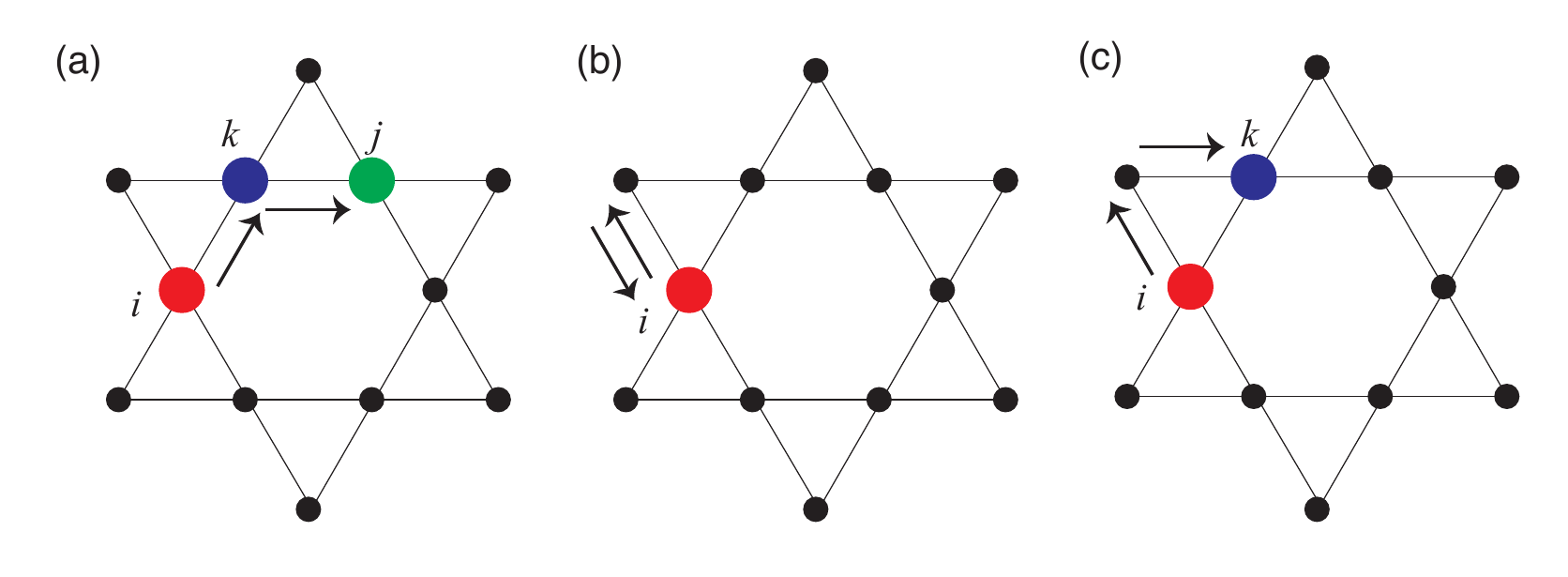}
\caption{
Three possibilities for two NN moves. Here we show an example on a kagome lattice.
(a) Starting from site $i$, one reaches site $j$, which is two Manhattan distance away from $i$.
(b) If one goes through a NN bond and goes back exactly along the same bond in the second move,
it ends up coming back to the original site $i$.
(c) It is also possible that, as a result of two NN moves, one reaches a NN site of $i$, namely, $k$. 
} 
\label{fig1}
\end{figure}

In this section, we outline our method to tune the flat band energy, and  
clarify the condition for this method to work. 
We consider a tight-binding Hamiltonian for spinless fermions with NN hoppings: 
\begin{equation}
\mathcal{H}_1 = t_1 \sum_{\langle i,j\rangle_{\rm NN}} c^{\dagger}_i c_j + c^{\dagger}_j c_i,
\end{equation}
where $t_1$ denotes the NN hopping integral. 
For future use, we also write down this Hamiltonian by using the incident
matrix of the lattice:
\begin{equation}
\mathcal{H}_1 = t_1 \sum_{i,j }c_i^\dagger [\hat{h}_1]_{i,j} c_j,
\end{equation}
where the incident matrix, $\hat{h}_1$, satisfies $[\hat{h}_1]_{i,j} = 1$ if $\langle i,j\rangle \in $ NN, and otherwise $[\hat{h}_1]_{i,j} = 0$.
We note that, throughout this paper, $h$ denotes dimensionless matrices in either real or momentum spaces. 
Their eigenvalues are denoted by $\lambda$, while energy eigenvalues of Hamiltonians are denoted by $\varepsilon$.

Suppose that the model is defined on a lattice with the number of sublattices $N_{\rm sub}$,
and that all sublattices have the same coordination number $z$.
We label sites as $i = (n,\alpha)$ where $n$ denotes a label of a unit cell and 
$\alpha$ labels the sublattice. 
By performing the Fourier transformation, we obtain 
\begin{equation}
\mathcal{H}_1 = t_1  \sum_{\bm{k},\alpha,\beta } c^\dagger_{\bm{k},\alpha} [h_1 (\bm{k})]_{\alpha \beta} c_{\bm{k}, \beta },
\end{equation}
where $c_{\bm{k},\alpha } = \sum_{n} c_{n,\alpha} e^{-i \bm{k} \cdot (\bm{R}_n + \bm{r}_\alpha) }$; 
$\bm{R}_n$ is the position of the unit cell 
and $\bm{r}_{\alpha}$ is the position of the sublattice $\alpha$ inside the unit cell. 

Let us assume that $h_1(\bm{k})$ has $N_f (< N_{\rm sub})$ flat bands and $N_{\rm sub} -N_f$ dispersive bands. 
We label wave functions of flat [dispersive] bands at each $\bm{k}$ as $\psi^{\rm (f) }_p (\bm{k})$ [$\psi^{\rm (d) }_q (\bm{k})$] 
and its eigenvalue $\lambda_p$ [$\lambda_q(\bm{k})$] with $p = 1, \cdots N_f$ [$q = 1, \cdots N_{\rm} - N_f$]. 
The corresponding eigenvalue equations are written as
\begin{equation}
h_1 (\bm{k}) \psi^{\rm (f) }_p (\bm{k}) = \lambda_p \psi^{\rm (f) }_p (\bm{k}),  \label{eq:eigenflat1}
\end{equation}
and 
\begin{equation}
h_1 (\bm{k}) \psi^{\rm (d) }_q (\bm{k}) = \lambda_q(\bm{k}) \psi^{\rm (d) }_q (\bm{k}).  \label{eq:eigendisp1}
\end{equation}

Under this setup, we now introduce our main idea for tuning the flat-band energy, that is,
we utilize the fact that if $\psi (\bm{k})$ is an eigenfunction of 
$h_1 (\bm{k})$, so it is of $[h_1 (\bm{k})]^{m}$
for $m$ being arbitrary positive integer.
More generally, {\it $\psi (\bm{k})$ is an eigenfunction for any polynomial of $h_1 (\bm{k})$.}
For instance, if we consider a generic quadratic form of $h_1 (\bm{k})$ with real coefficients $a,b$ and $c$, 
we obtain the eigenvalue equations as
\begin{eqnarray}
& \{a [h_1 (\bm{k})]^2  + b h_1(\bm{k}) +c \hat{I}_{N_{\rm sub}}
\}\psi^{\rm (f) }_p (\bm{k})  \nonumber \\
& =  [a (\lambda_p)^2   + b\lambda_p + c ]\psi^{\rm (f) }_p (\bm{k}),  \label{eq:eigenflat2}
\end{eqnarray}
and 
\begin{eqnarray}
& \{a [h_1 (\bm{k})]^2  + b h_1(\bm{k}) +c \hat{I}_{N_{\rm sub}}
\} \psi^{\rm (d) }_q (\bm{k})  \nonumber \\
& = \{a [\lambda_q (\bm{k}) ]^2   + b [\lambda_q(\bm{k})] + c \} \psi^{\rm (d) }_q (\bm{k}),  \label{eq:eigendisp2}
\end{eqnarray}
where $ \hat{I}_{N_{\rm sub}}$ denotes $N_{\rm sub}\times N_{\rm sub}$ identity matrix.
Then, the new eigenvalues $a (\lambda_p)^2   + b\lambda_p + c$ and $a [\lambda_q (\bm{k}) ]^2   + b [\lambda_q(\bm{k})] + c$
can intersect on some lines or surfaces in the Brillouin zone, even if the original eigenvalues, $\lambda_p$ and $\lambda_q(\bm{k})$, do not. 

How can we implement a polynomial of $h_1 (\bm{k})$ in the tight-binding models? 
To see this, let us come back to the real-space representation, in which the square and higher powers of $\hat{h}_1$ have a simple interpretation. $[\hat{h}_1^2]_{ij}=\sum_k[\hat{h}_1]_{ik}[\hat{h}_1]_{kj}$ is finite, only if there is a site $k$ neighboring both site $i$ and $j$, i.e., if site $j$ can be reached from site $i$ by two successive NN hoppings. Generalizing it, the $m$-th power of $\hat{h}_1$, $\hat{h}_1^m$, has finite matrix element, $[\hat{h}_1^m]_{ij}$, only if sites $i$ and $j$ are $m$ NN hoppings away. To discuss the structure of $\hat{h}_1^m$ in a systematic way, it is convenient to introduce Manhattan distance of the graph.

The Manhattan distance between two sites, say $i$ and $j$, is defined as the minimum number of NN bonds one has to go through when 
moving from $i$ to $j$ along the bonds. 
For instance, if the Manhattan distance between $i$ and $j$ is two, it means that there exists a site 
$k$ such that both $i$ and $j$ are connected to $k$ and $j$ is not the NN of $i$ [Fig. \ref{fig1}(a)]. 

At first sight, the above argument implies that $\hat{h}_1^2$ is proportional to the incident matrix of the Manhattan distance two, i.e., $[\hat{h}_1^2]_{ij}$ is finite only if $i$ and $j$ are separated by the Manhattan distance of two.
Indeed, if $i$ and $j$ are two Manhattan distances away, we have a finite matrix element, $[\hat{h}_1^2]_{i,j} = x$,
where $x$ is the number of sites neighboring both $i$ and $j$.  
However, we have to keep in mind that, if you move twice along NN bonds, there are two other possibilities,
other than reaching a site of two Manhattan distances away.
The first possibility is coming back to the original site, which occurs when going through the same bond twice [Fig. \ref{fig1}(b)].
The second possibility is reaching the NN site [Fig. \ref{fig1}(c)]. 
Let us assume that, for every NN pair, say $i$ and $j$, 
there are $y$ distinct paths going from $i$ to $j$ with passing two NN bonds.
In other words, there exist sites $\ell_1, \cdots \ell_y  \neq i,j$, 
such that $\langle i,\ell_n \rangle \in$ NN and $\langle j,\ell_n \rangle \in$ NN for $n= 1, \cdots y$. 

Under this assumption, we obtain the incident
matrix for a Manhattan distance two as
 \begin{equation}
[(\hat{h}_1)^2]_{i,j} = x[\hat{h}_2]_{i,j} + y [\hat{h}_1]_{i,j}+ z \delta_{i,j},
\end{equation}
where $\delta_{i,j}$ is the Kronecker delta. 
Alternatively, in the momentum space representation, we obtain
\begin{equation}
[h_1(\bm{k})]^2 = x h_2(\bm{k}) + y \hat{h}_1(\bm{k}) +z\hat{I}_{N_{\rm sub}},
\end{equation}
where $h_2(\bm{k})$ is a (dimensionless) hopping matrix for ``second-neighbor" hoppings. 
Therefore, if we introduce the second-neighbor hoppings with a transfer integral $t_2$, 
we obtain the quadratic form of $h_1(\bm{k})$ as 
\begin{widetext}
\begin{eqnarray}
\mathcal{H} =  \sum_{\bm{k},\alpha,\beta}  c^{\dagger}_{\bm{k},\alpha} [t_1 h_1(\bm{k}) + t_2 h_2(\bm{k}) ]_{\alpha \beta} c_{\bm{k},\beta}
 =  \sum_{\bm{k},\alpha,\beta}  c^{\dagger}_{\bm{k},\alpha} \left\{ t_2\frac{1}{x} [h_1(\bm{k})]^2 + (t_1- t_2\frac{y}{x}) h_1(\bm{k}) - t_2\frac{z}{x}\hat{I}_{N_{\rm sub}}  \right\}_{\alpha \beta} c_{\bm{k},\beta}. \label{}
 \nonumber \\ \label{eq:ham_general}
\end{eqnarray}
\end{widetext}
Consequently, the eigenenergies of this Hamiltonian are
$f(\lambda_{p/q})$ with $f(\lambda) =t_2 \frac{1}{x}\lambda^2 +(t_1 - t_2\frac{y}{x})\lambda - t_2\frac{z}{x}$. 

In the next two sections, we demonstrate how this idea works through the analyses of specific models. 
We first show canonical examples in kagome and pyrochlore models in Sec. \ref{sec:result1}. 
In these lattices, the aforementioned lattice parameters, e.g., $x$, $y$ and $z$, are sublattice-independent,
thus these lattices are ``homogeneous". 
In Sec. \ref{sec:result2}, we discuss the applications to breathing lattices and a class of Lieb lattices,
in which existence of inequivalent sites or bonds modifies the simple polynomial expression mentioned above.

Before closing this section, we remark that 
higher-order polynomials of $h_1(\bm{k})$ can be obtained by introducing 
the ``farther-neighbor" Manhattan-distance-dependent hoppings. 
However, in light of material realization, short-ranged hoppings are favorable. 
Moreover, remarkable tunability of band structure is available, even within the model with second Manhattan distance, as we will show below.

\section{Canonical examples: kagome and pyrocholore lattices \label{sec:result1}}
We apply the idea we discussed in the previous section  
to the kagome and pyrochlore lattices, which are the textbook examples of flat-band models.
In a previous study, 
the authors investigated the band structures on these models in the context of magnetic mode analysis~\cite{Mizoguchi2018}.
In this paper, we discuss their band structures, focusing on the quantities relevant to electronic systems, such as the DOS.

\subsection{Kagome lattice \label{sec:kagome}}
\begin{figure*}
\centering
\includegraphics[width=\linewidth]{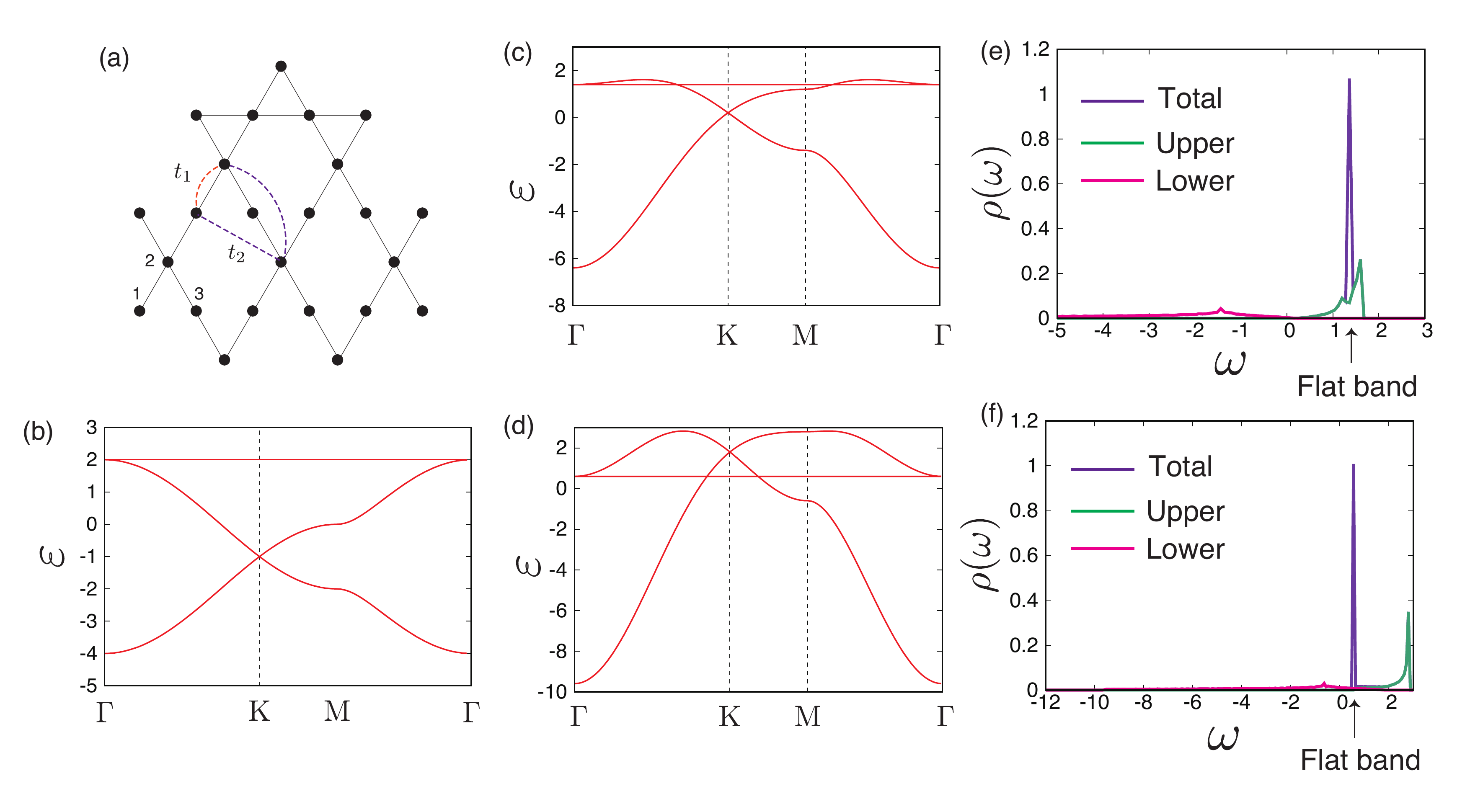}
\caption{
(a) A kagome lattice.
Orange (purple) dashed lines denote the hopping processes with the hopping integral $t_1$ ($t_2$).
The sublattices are denoted by 1, 2 and 3.
The band structures for (b) $(t_1,t_2) = (-1,0) $, (c) $ (-1,-0.3) $, and (d) $ (-1,-0.7)$.
The high-symmetry points in the Brillouin zone are given by 
$\Gamma = (0,0)$, K$=(\frac{4\pi}{3},0)$ and M$=(\pi,\frac{\pi}{\sqrt{3}})$.
The DOS for (e) $ (-1,-0.3) $, and (f) $ (-1,-0.7)$.
} 
\label{fig:kagome}
\end{figure*}
We first show the results for a kagome lattice.
We take the lattice vectors as $\bm{a}_1^{\rm(K)} = (1,0)$, $\bm{a}_2^{\rm(K)} = \left(\frac{1}{2}, \frac{\sqrt{3}}{2} \right)$
and the sublattices' coordinates as $\bm{r}_1^{\rm(K)} = (0,0)$, $\bm{r}_2^{\rm(K)} = \left(\frac{1}{4}, \frac{\sqrt{3}}{4} \right)$,
$\bm{r}_3^{\rm(K)} = \left(\frac{1}{2},0 \right)$. 
Then, the NN hopping matrix in the momentum space is given by
\begin{equation}
h^{\rm (K)} _1 (\bm{k}) = \left(
\begin{array}{ccc}
0 & h_{12}^{\rm (K,1)} (\bm{k})& h_{13}^{\rm (K,1)} (\bm{k})\\
h_{12}^{\rm (K,1)} (\bm{k})&0 & h_{23}^{\rm (K,1)} (\bm{k}) \\
h_{13}^{\rm (K,1)} (\bm{k}) & h_{23}^{\rm (K,1)} (\bm{k})& 0\\
\end{array}
\right),
\end{equation}
with $h_{12}^{\rm (K,1)} (\bm{k}) =2 \cos \left( \frac{k_x + \sqrt{3} k_y}{4}\right)$,
$h_{13}^{\rm (K,1)} (\bm{k}) =2 \cos \left( \frac{k_x }{2}\right)$, 
and $h_{23}^{\rm (K,1)} (\bm{k}) =2 \cos \left(\frac{k_x - \sqrt{3} k_y}{4} \right)$.
Due to the nature of a line graph, $h^{\rm (K)}_1(\bm{k})$
has a $\bm{k}$-independent eigenvalue $\lambda^{\rm (K,f)} =-2$.
The other two bands are dispersive and their dispersion relations are given as 
\begin{widetext}
\begin{eqnarray}
\lambda^{\rm (K,d)}_{1} (\bm{k}) = 1+ \sqrt{2\left[ \cos k_x + \cos\left( \frac{k_x + \sqrt{3}k_y}{2}\right) + \cos\left( \frac{k_x - \sqrt{3}k_y}{2}\right) \right] + 3}, 
\nonumber \\
\end{eqnarray}
\begin{eqnarray}
\lambda^{\rm (K,d)}_{2} (\bm{k}) = 1- \sqrt{2\left[ \cos k_x + \cos\left( \frac{k_x + \sqrt{3}k_y}{2}\right) + \cos\left( \frac{k_x - \sqrt{3}k_y}{2}\right) \right] + 3}.\nonumber \\
\end{eqnarray}
\end{widetext}
The corresponding eigenfunctions are given by
\begin{widetext}
\begin{eqnarray}
\psi^{\rm (K,f)} (\bm{k}) = \frac{1}{\mathcal{N}^{\rm (K,f)} (\bm{k})} \left[ \sin \left( \varphi_2 (\bm{k}) - \varphi_3 (\bm{k}) \right), \sin \left( \varphi_3 (\bm{k}) - \varphi_1 (\bm{k}) \right), \sin \left( \varphi_1 (\bm{k}) - \varphi_2 (\bm{k}) \right) \right]^{\mathrm{T}},
\nonumber \\
\end{eqnarray}
\begin{eqnarray}
\psi^{\rm (K,d)}_1 (\bm{k}) =  \frac{1}{\mathcal{N}_1^{\rm (K,d) }(\bm{k})} \left[ \cos \left(\varphi_1 (\bm{k})+ \theta (\bm{k}) \right), \cos \left(\varphi_2 (\bm{k}) + \theta (\bm{k})\right), \cos \left(  \varphi_3 (\bm{k}) + \theta (\bm{k}) \right)  \right]^{\mathrm{T}}, \nonumber \\
\end{eqnarray}
and 
\begin{equation}
\psi^{\rm (K,d)}_2 (\bm{k}) 
=\frac{1}{\mathcal{N}_2^{\rm (K,d)} (\bm{k})}  \left[ \sin \left( \varphi_1 (\bm{k})+ \theta (\bm{k})\right), \sin \left(\varphi_2 (\bm{k})+  \theta (\bm{k}) \right), \sin  \left( \varphi_3 (\bm{k})+ \theta (\bm{k})\right) \right]^{\mathrm{T}}, 
\end{equation}
\end{widetext}
where 
$\mathcal{N}^{\rm (K,f)} (\bm{k})$, $\mathcal{N}_1^{\rm (K,d) }(\bm{k})$ and $\mathcal{N}_2^{\rm (K,d) }(\bm{k})$ are the normalization factors,
$\varphi_1(\bm{k}) = \frac{k_x}{4} + \frac{k_y}{4\sqrt{3}}$, $\varphi_2(\bm{k}) =- \frac{k_y}{2\sqrt{3}}$, $\varphi_3(\bm{k}) =  - \frac{k_x}{4} + \frac{k_y}{4\sqrt{3}}$, 
and $\theta (\bm{k} )  = \frac{1}{2} \mathrm{arg} \left[ e^{i\frac{k_y}{\sqrt{3}} } + 2 \cos \frac{k_x}{2} e^{-i \frac{k_y}{2\sqrt{3} } }\right]$
are the phase factors arising from the geometry of the lattice. 
We show the band structure of the NN Hamiltonian with $t_1 = -1$ in Fig \ref{fig:kagome}(b).

Now, let us tune the flat-band energy. 
To this end, we introduce the second-neighbor hoppings: 
\begin{equation}
h^{\rm (K)}_2 (\bm{k}) = \left(
\begin{array}{ccc}
h_{11}^{\rm (K,2)} (\bm{k})&h_{12}^{\rm (K,2)} (\bm{k})& h_{13}^{\rm (K,2)} (\bm{k}) \\
h_{12}^{\rm (K,2)} (\bm{k})&h_{22}^{\rm (K,2)} (\bm{k}) &h_{23}^{\rm (K,2)} (\bm{k})\\
h_{13}^{\rm (K,2)} (\bm{k})&h_{23}^{\rm (K,2)} (\bm{k}) &h_{33}^{\rm (K,2)} (\bm{k}) \\
\end{array}
\right),
\end{equation}
where $h_{11}^{\rm (K,2)} (\bm{k}) = 2 \left( \cos k_x +  \cos \frac{k_x + \sqrt{3}k_y}{2}\right)$, 
$h_{12}^{\rm (K,2)} (\bm{k}) = 2 \cos \left( \frac{3 k_x -\sqrt{3} k_y}{4}  \right)$,
$h_{13}^{\rm (K,2)} (\bm{k}) = 2 \cos \left( \cos \frac{ \sqrt{3}k_y}{2} \right)$,
$h_{22}^{\rm (K,2)} (\bm{k}) = 2 \left( \cos \frac{k_x + \sqrt{3}k_y}{2} +  \cos \frac{k_x - \sqrt{3}k_y}{2}\right)$,
$h_{23}^{\rm (K,2)} (\bm{k}) = 2 \cos \left( \frac{3 k_x + \sqrt{3} k_y}{4}  \right)$,
and 
$h_{33}^{\rm (K,2)} (\bm{k}) = 2 \left( \cos k_x +  \cos \frac{k_x -  \sqrt{3}k_y}{2}\right)$. 
Constructed as such, $h^{\rm (K)}_2 (\bm{k})$ is expressed by a quadratic form of $h^{\rm (K)}_1 (\bm{k})$, as we have seen in the previous section. 
Indeed, one can show that 
$h^{\rm (K)}_2 (\bm{k})$ can be written by using $h^{\rm (K)}_1 (\bm{k})$ as
\begin{equation}
h^{\rm (K)}_2 (\bm{k}) = \left[ h^{\rm (K,1)}_1(\bm{k}) \right]^2 - h^{\rm (K,1)}_1 (\bm{k})- 4\hat{I}_{3}.  \label{eq:kagome_poly}
\end{equation}
since $(x,y,z) = (1,1,4)$ for a kagome lattice. 

Then, let us consider the Hamiltonian 
\begin{equation}
\mathcal{H}^{\rm (K)} = \sum_{ \bm{k} } \hat{c}_{\bm{k}} ^{\dagger} \left[t_1 h_1^{\rm(K)} (\bm{k}) + t_2 h_2^{\rm(K)} (\bm{k})\right]. 
\end{equation}
The band dispersion of $\mathcal{H}^{\rm (K)}$ is obtained by using Eq. (\ref{eq:kagome_poly}) as 
\begin{equation}
\varepsilon^{\rm (K,f)} =- 2(t_1 -t_2), 
 \end{equation}
\begin{eqnarray}
\varepsilon^{\rm (K,d)}_1 (\bm{k}) =  & f(\lambda^{\rm (K,d)}_1(\bm{k}))  \nonumber \\
\equiv  & t_2 [\lambda^{\rm (K,d)}_1(\bm{k})]^2  + (t_1-t_2) \lambda^{\rm (K,d)}_1(\bm{k}) - 4t_2, \nonumber \\
 \end{eqnarray}
 \begin{eqnarray}
\varepsilon^{\rm (K,d)}_2 (\bm{k}) = & f(\lambda^{\rm (K,d)}_2(\bm{k}))  \nonumber \\
\equiv  & t_2 [\lambda^{\rm (K,d)}_2(\bm{k})]^2  + (t_1-t_2) \lambda^{\rm (K,d)}_2(\bm{k}) - 4t_2.\nonumber \\
 \end{eqnarray}
Notice that, although the flat and dispersive bands touch at $t_2=0$~\cite{Bergman2008,Rhim2018}, 
the intersection among these band does not occur as soon as infinitesimal $t_2$ is introduced.
Indeed, in the previous study, the authors have shown that this occurs when $t_1$ and $t_2$ have the same sign and they satisfy 
$|t_2| > |t_1|/5$~\cite{Mizoguchi2018}. 

The intersection of bands leads to the divergence of partial DOS contributed from dispersive bands.
As the introduction of $t_2$, the partial DOS, $\rho^{0}_q(\varepsilon)$, contributed from the original dispersive bands, $\lambda^{\rm (K,d)}_q(\bm{k})$, are deformed as
\begin{eqnarray}
\rho_q(\varepsilon) = \frac{1}{|f'(\varepsilon)|}\rho^{0}_q(f^{-1}(\varepsilon)).
\end{eqnarray}

As an example of the band intersection, we plot a band structure for $(t_1,t_2) = (-1,-0.3)$ in Fig. \ref{fig:kagome}(c).
We also show the DOS, $\rho (\omega)$ for the same parameter in Fig. \ref{fig:kagome}(e). 
Here the DOS is computed numerically as 
\begin{eqnarray}
 \rho (\omega)
= & \frac{1}{N_{\mathrm{m}} }\sum_{\bm{k},n} \Theta \left( \varepsilon_n(\bm{k}) - \left(\omega -\frac{\Delta \omega}{2}\right) \right) \nonumber \\
\times & \Theta \left( \left( \omega + \frac{\Delta \omega}{2}\right) -\varepsilon_n(\bm{k}) \right), \nonumber \\
\end{eqnarray}
where 
$n$ is the label of bands, $\Delta \omega$
is a unit of discretized energy set as $0.08$,
$N_{\mathrm{m}}$ is a number of mesh in the momentum space
set as $N_{\mathrm{m}} = 200 \times 200$,
and 
$\Theta \left( x \right)$ is a Heaviside step function. 

We see that, other than the contribution from the flat band, there is large DOS at the band top.
This is due to the fact that the band maxima form a line in the two-dimensional Brillouin zone, rather than discrete points, 
meaning that it has a sub-extensive degeneracy~\cite{Mizoguchi2018}.
This causes the divergence of the DOS at the band top.
In fact, from the relation between the original and modified dispersive bands, our method generally leads to the $d-1$ dimensional degenerate surface at the band top, giving rise to the strongly divergent DOS proportional to $\varepsilon^{-1/2}$, irrespective of the system dimension, $d$.
Since the flat band is relatively close to the band top, 
the obtained band structure is potentially suitable for obtaining 
high-$T_c$ superconductivity due to the interband scattering. 

For comparison, we also show the results for $(t_1,t_2) = (-1,-0.7)$ in Figs. \ref{fig:kagome}(d) and  \ref{fig:kagome}(f).
Although the penetration of flat band occurs as well, the DOS of dispersive band is more or less small near the flat band.
This is due to the fact that the flat-band energy is far from the band top, where the upper dispersive band has a large DOS. 

 \subsection{Pyrochlore lattice}
\begin{figure}
\centering
\includegraphics[width=\linewidth]{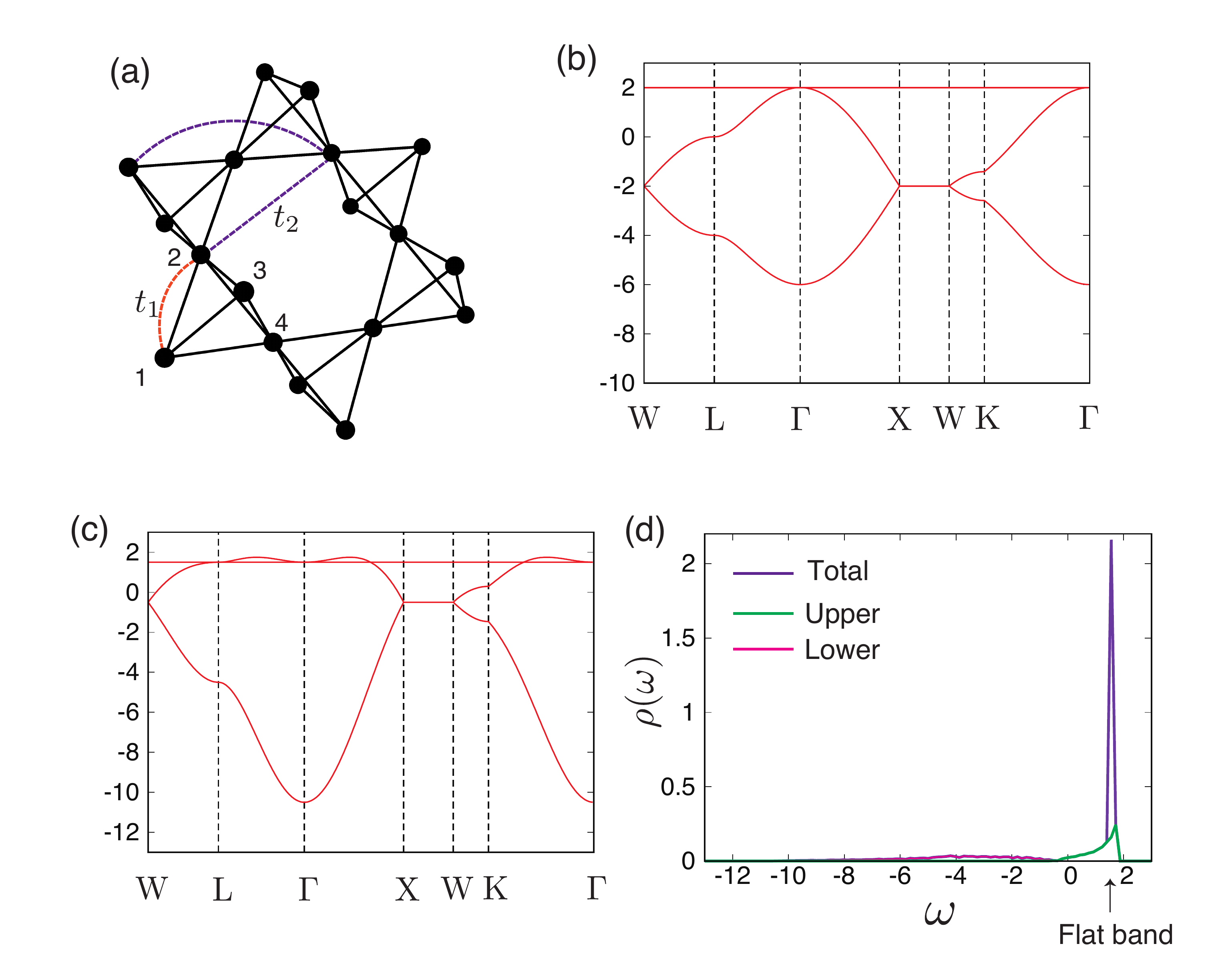}
\caption{
(a) A pyrochlore lattice.
Orange (purple) dashed lines denote the hopping processes with the hopping integral $t_1$ ($t_2$).
The sublattices are denoted by 1, 2 3 and 4.
The band structures for (b) $(t_1,t_2) = (-1,0) $, and (c) $ (-1,-0.25) $.
The high-symmetry points in the Brillouin zone are given by 
$\Gamma = (0,0,0)$,
X$=(0,0,2\pi)$,
K$=(\frac{3\pi}{2},0,\frac{3\pi}{2})$,
W$=(\pi,0,2\pi)$,
and
L$=(\pi,\pi,\pi)$.
(d) The DOS for $(t_1,t_2) = (-1,-0.25)$. 
Purple line is from all four band.
Green and magenta lines are from upper and lower dispersive bands, respectively. 
} 
\label{fig:pyrochlore}
\end{figure}
We can apply the same method to a pyrochlore lattice.
The lattice vectors are $\bm{a}_1^{\rm (P)} = (0,1/2,1/2)$, $\bm{a}_2^{\rm (P)} = (1/2,0,1/2)$, and $\bm{a}_3^{\rm (P)} = (1/2,1/2,0)$.
The coordinates of sublattices are $\bm{r}_1^{\rm (P)} = (0,0,0)$, $\bm{r}_2^{\rm (P)} = (0,1/4,1/4)$, $\bm{r}_3^{\rm (P)} = (1/4,0,1/4)$, and $\bm{r}_4^{\rm (P)} = (1/4,1/4,0)$.

The NN Hamiltonian in the momentum space is then given by
\begin{equation}
h^{\rm(P)}_1 (\bm{k}) =\left(
\begin{array}{cccc}
0 & h^{\rm(P,1)}_{12} (\bm{k})&h^{\rm(P,1)}_{13} (\bm{k})&h^{\rm(P,1)}_{14} (\bm{k}) \\
h^{\rm(P,1)}_{12} (\bm{k}) &0 & h^{\rm(P,1)}_{23} (\bm{k}) &h^{\rm(P,1)}_{24} (\bm{k}) \\
h^{\rm(P,1)}_{13} (\bm{k})& h^{\rm(P,1)}_{23} (\bm{k})& 0 &   h^{\rm(P,1)}_{34} (\bm{k})\\
h^{\rm(P,1)}_{14} (\bm{k})& h^{\rm(P,1)}_{24} (\bm{k}) &h^{\rm(P,1)}_{34} (\bm{k}) & 0 \\
\end{array}
\right)
\end{equation}
with $h^{\rm(P)}_{12} (\bm{k}) = 2\cos \left(\frac{k_y + k_z}{4}\right)$,
$h^{\rm(P)}_{13} (\bm{k}) = 2\cos \left(\frac{k_x + k_z}{4}\right)$,
$h^{\rm(P)}_{14} (\bm{k}) = 2\cos \left(\frac{k_x + k_y}{4}\right)$,
$h^{\rm(P)}_{23} (\bm{k}) = 2\cos \left(\frac{k_x - k_y}{4}\right)$,
$h^{\rm(P)}_{24} (\bm{k}) = 2\cos \left(\frac{k_x - k_z}{4}\right)$,
and $h^{\rm(P)}_{34} (\bm{k}) = 2\cos \left(\frac{k_y - k_z}{4}\right)$.
$h^{\rm(P)}_1 (\bm{k})$ has two flat eigenvalues, $\lambda_{1}^{\rm (P)} = \lambda_{2}^{\rm (P)} = -2$,
and the other two eigenvalues are 
\begin{equation}
\lambda^{\mathrm{(P)}}_1 (\bm{k}) =  2 +  \sqrt{4 + F^{\rm(P)}(\bm{k}) },
\end{equation}
\begin{equation}
\lambda^{\mathrm{(P)}}_2 (\bm{k}) = 2 -\sqrt{4 + F^{\rm(P)}(\bm{k})},
\end{equation}
with 
\begin{widetext}
\begin{eqnarray}
F^{\rm(P)}(\bm{k}) \equiv 2 \left[ \cos \left( \frac{k_x + k_y}{2}\right) + \cos \left( \frac{k_y + k_z}{2}\right) +\cos \left( \frac{k_z + k_x}{2}\right) 
+\cos \left( \frac{k_x - k_y}{2}\right) + \cos \left( \frac{k_y - k_z}{2}\right) +\cos \left( \frac{k_z - k_x}{2}\right) \right]. \nonumber \\
\end{eqnarray}
\end{widetext}

As we did for kagome, 
we introduce the second-neighbor hoppings as
\begin{equation}
h_2 (\bm{k}) =  \left(
\begin{array}{cccc}
h^{\rm (P,2) }_{11} (\bm{k}) &h^{\rm (P,2)}_{12} (\bm{k})  & h^{\rm (P,2)}_{13} (\bm{k})&  h^{\rm (P,2)}_{14} (\bm{k}) \\
h^{\rm (P,2) }_{12} (\bm{k})&h^{\rm (P,2)}_{22} (\bm{k}) & h^{\rm (P,2)}_{23} (\bm{k})  &h^{\rm (P,2)}_{24} (\bm{k}) \\
h^{\rm (P,2) }_{13} (\bm{k}) &h^{\rm (P,2) }_{23} (\bm{k})  &h^{\rm (P,2) }_{33} (\bm{k}) &h^{\rm (P,2) }_{34} (\bm{k}) \\
h^{\rm (P,2) }_{14} (\bm{k})  &  h^{\rm (P,2) }_{24} (\bm{k})&h^{\rm (P,2) }_{34} (\bm{k}) & h^{\rm (P,2) }_{44} (\bm{k})\\
\end{array}
\right),
\end{equation}
with $h^{\rm (P,2) }_{11} (\bm{k}) = 2 \left[\cos\left( \frac{k_x +k_y}{2} \right)+ \cos\left( \frac{k_z +k_x}{2} \right)+ \cos\left( \frac{k_y +k_z}{2} \right) \right]$, 
$h^{\rm (P,2) }_{12} (\bm{k}) = 4  \cos\left(\frac{k_x}{2} \right) \cos\left( \frac{k_y -k_z}{4}\right)$, 
$h^{\rm (P,2) }_{13} (\bm{k}) =  4  \cos\left(\frac{k_y}{2} \right) \cos\left( \frac{k_x -k_z}{4}\right)$, 
$h^{\rm (P,2) }_{14} (\bm{k}) = 4  \cos\left(\frac{k_z}{2} \right) \cos\left( \frac{k_x-  k_y}{4}\right)$,
$h^{\rm (P,2)}_{22} (\bm{k}) = 2 \left[ \cos\left( \frac{k_x - k_y}{2} \right)+ \cos\left( \frac{k_z - k_x}{2} \right)+ \cos\left( \frac{k_y +k_z}{2} \right) \right] $,
$h^{\rm (P,2) }_{23} (\bm{k}) = 4  \cos\left(\frac{k_z}{2} \right) \cos\left( \frac{k_x +  k_y}{4}\right)$,
$h^{\rm (P,2) }_{24} (\bm{k}) = 4  \cos\left(\frac{k_y}{2} \right) \cos\left( \frac{k_x +  k_z}{4}\right)$,
$h^{\rm (P,2)}_{33} (\bm{k}) = 2 \left[ \cos\left( \frac{k_x - k_y}{2} \right)+ \cos\left( \frac{k_z + k_x}{2} \right)+ \cos\left( \frac{k_y - k_z}{2} \right) \right] $,
$h^{\rm (P,2) }_{34} (\bm{k}) = 4  \cos\left(\frac{k_x}{2} \right) \cos\left( \frac{k_y +  k_z}{4}\right)$,
and $h^{\rm (P,2)}_{44} (\bm{k}) = 2 \left[ \cos\left( \frac{k_x + k_y}{2} \right)+ \cos\left( \frac{k_z - k_x}{2} \right)+ \cos\left( \frac{k_y - k_z}{2} \right) \right] $.

Since the lattice parameters are given as $(x,y,z) = (1,2,6)$, $h_2 (\bm{k})$ satisfies
\begin{equation}
h^{\rm (P)}_2 (\bm{k}) = \left[ h^{\rm (P,1)}_1(\bm{k}) \right]^2 - 2 h^{\rm (P,1)}_1(\bm{k}) - 6\hat{I}_{4}.  \label{eq:pyro_poly}
\end{equation}

Now let us consider the Hamiltonian 
\begin{equation}
\mathcal{H}^{\rm (P)} = \sum_{ \bm{k} } \hat{c}_{\bm{k}} ^{\dagger} \left[t_1 h_1^{\rm(P)} (\bm{k}) + t_2 h_2^{\rm(P)} (\bm{k})\right].
\end{equation}
Then, if $t_1$ and $t_2$ have the same sign and $|t_2|/ |t_1| > 1/6$, the flat bands penetrate the dispersive band~\cite{Mizoguchi2018}. 

We show the band structure and DOS for $(t_1,t_2) = (-1,-0.25)$ in Figs. \ref{fig:pyrochlore}(c) and \ref{fig:pyrochlore}(d), respectively.
Here we use $32 \times 32 \times 32$ meshes in the Brillouin zone for the summation over $\bm{k}$.
Again, the upper dispersive band has relatively large DOS near the band top, which is penetrated by the flat band. 

\section{Extensions to inhomogeneous models \label{sec:result2}}
Our method is also applicable to the models with site/bond inequivalency.
We first consider ``breathing" lattices of kagome and pyrochlore, 
where the bond inhomogeneity is introduced to the original kagome or pyrochlore lattices. 
These lattices are recently of interests particularly in the contest of higher-order topological insulators~\cite{Hatsugai2011,Xu2017,Kunst2018,Romhanyi2018,Ezawa2018}, as well as frustrated magnetism~\cite{Okamoto2013,Benton2015,Shaffer2017,Orain2017,Tsunetugu2017,Essafi2017}. 
We also consider non-line-graph lattices, such as a Lieb lattice~\cite{Lieb1989} and a dice lattice~\cite{Sutherland1986},
as examples of site-inhomogeneous lattices. 

\subsection{Breathing kagome and prochlore lattices}
\begin{figure}[t]
\centering
\includegraphics[width=\linewidth]{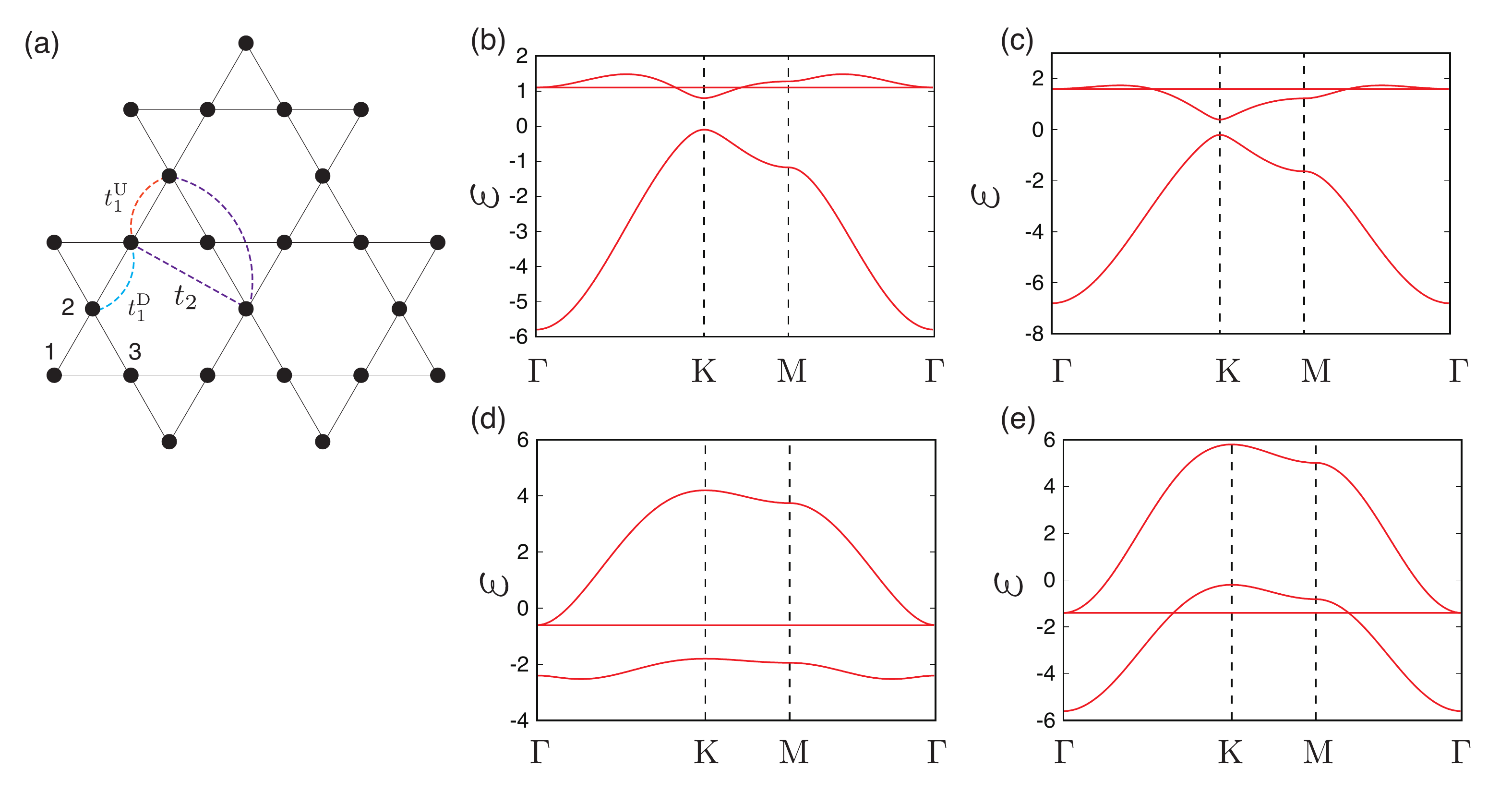}
\caption{
(a) Hoppings for a breathing kagome lattice. Orange, light-blue and purple lines denote $\tu$, $\td$ and $t_2$, respectively.
Band structures for $(\tu,\td,t_2)$ equals to (b) $(-1,-0.7,-0.3)$, (c) $(-1,-1.2,-0.3)$, (d)  $(-1,1,-0.3)$, and (e) $(-1,1,-0.7)$. 
} 
\label{fig:bkagome}
\end{figure}

\begin{figure}[t]
\centering
\includegraphics[width=\linewidth]{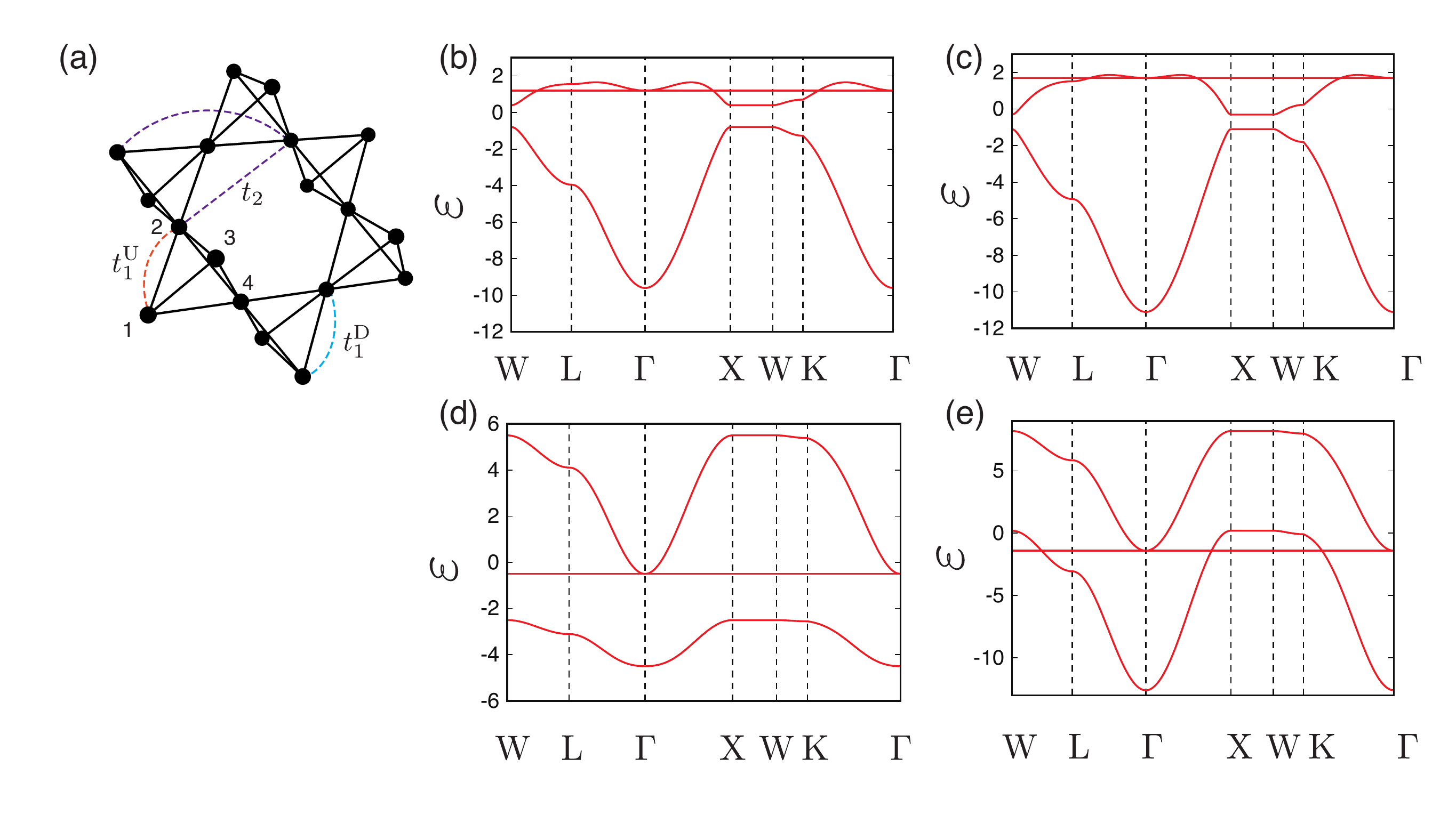}
\caption{
(a) Hoppings for a breathing pyrochlore lattice. Orange, light-blue and purple lines denote $\tu$, $\td$ and $t_2$, respectively.
Band structures for $(\tu,\td,t_2)$ equals to (b) $(-1,-0.7,-0.25)$, (c) $(-1,-1.2,-0.25)$, (d)  $(-1,1,-0.25)$, and (e) $(-1,1,-0.7)$. 
} 
\label{fig:bpyrochlore}
\end{figure}

In the breathing kagome (pyrochlore) lattice, the transfer integrals are modulated 
from the original models in such a way 
that the transfer integral on upward triangles (tetrahedra), $t_1^{\rm U}$, is not equal to that on downward ones, $t_1^{\rm D}$.
Our method works even in breathing lattices despite the presence of bond inequivalency, because,
the eigenfunctions for a flat band does not change even if we introduce the breathing-type modulation~\cite{Hatsugai2011}. 

At the NN model, 
the ``position" of flat band(s) is sensitive to the relative sign between $t_1^{\rm U}$ and $t_1^{\rm D}$~\cite{Hatsugai2011, Essafi2017,Ezawa2018}. 
If these two have the same sign, the flat band resides in the band top or bottom.
If they are opposite, on the other hand, it is located in the middle of two dispersive bands, keeping touching points with
either upper or lower bands. 

When we introduce the second-neighbor hoppings, the flat-band penetration occurs in both cases for sufficiently large $|t_2|$, 
but in a quite different manner. 

First, let us see the case where both $t_1^{\rm U}$ and $t_1^{\rm D}$ have a negative sign. 
In this case, the flat band penetrates the upper band for both $|t_1^{\rm U}| > |t_1^{\rm D}|$ and $|t_1^{\rm U}| < |t_1^{\rm D}|$
[Figs. \ref{fig:bkagome}(b) and \ref{fig:bkagome}(c) for a breathing kagome, and Figs. \ref{fig:bpyrochlore}(b) and \ref{fig:bpyrochlore}(c) for a breathing pyrochlore],
as is in the case of the original kagome and pyrochlore lattices.

Next, we consider the case of opposite sign, in particular, the case with $t_1^{\rm U} = -1$ and $t_1^{\rm D}  =1$. 
In the absence of $t_2 (<0) $, the flat band intersects the line node
of the dispersive band at the $\Gamma$ point~\cite{Essafi2017}. 
This line node reminds us of a Dirac cone, however, the structure of the eigenfunction comprising this line node structure is rather 
close to the bosonic magnon mode associated with antiferromagnetic ordering, i.e., it is a fermionic realization of a Goldstone mode~\cite{Essafi2017}.
When we introduce small but finite $t_2$, 
we first see that the line node 
is gapped out, and the flat band stays 
touched with the upper dispersive band at the $\Gamma$ point  
[Fig. \ref{fig:bkagome}(d) for a breathing kagome, and Fig. \ref{fig:bpyrochlore}(d) for a breathing pyrochlore]. 
Upon increasing $|t_2|$, we see that the flat band penetrates the lower dispersive band, 
while retaining a band-touching point with the upper dispersive band 
[Fig. \ref{fig:bkagome}(e) for a breathing kagome, and Fig. \ref{fig:bpyrochlore}(e) for a breathing pyrochlore]. 

The evolution of the aforementioned band structures is tracked by the analytical formulas of the dispersion relations given in the appendix. 

 \subsection{Lieb lattice}
\begin{figure}
\centering
\includegraphics[width=\linewidth]{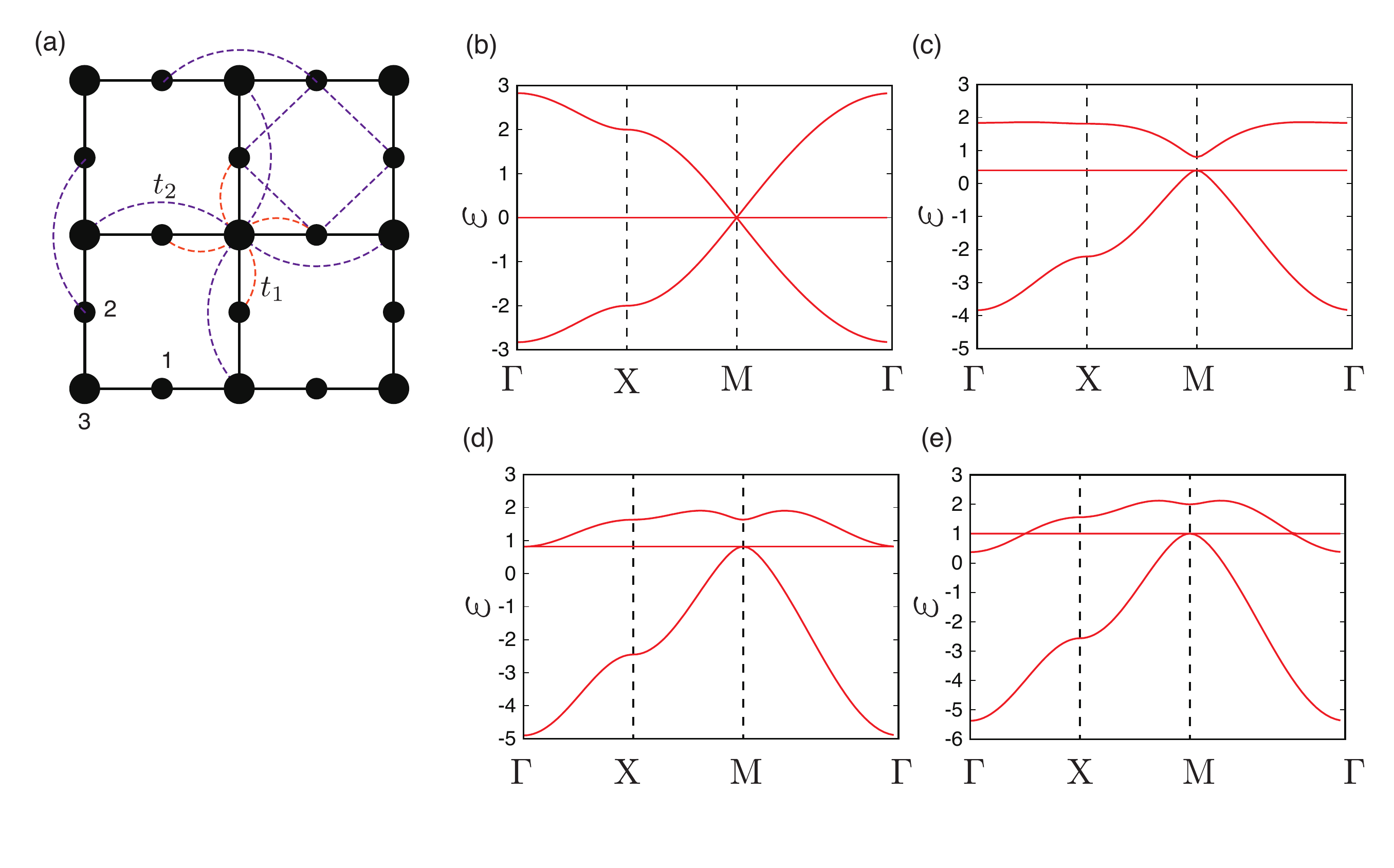}
\caption{
(a) A Lieb lattice.
Orange (purple) dashed lines denote the hopping processes with the hopping integral $t_1$ ($t_2$).
The sublattices are denotes by 1, 2 and 3.
The band structures for (b) $(t_1,t_2) = (-1,0) $, (c) $(t_1,t_2) = (-1,-0.2) $, (d) $(t_1,t_2) = (-1,-1/\sqrt{6})$, and (e) $(t_1,t_2) = (-1,-0.5)$.
The high-symmetry points in the Brillouin zone are given by 
$\Gamma = (0,0)$, X$=(\pi,0)$, and M$=(\pi,\pi)$.
} 
\label{fig:Lieb}
\end{figure}
In the following two subsections, we consider a class of Lieb lattices, as examples of site-inhomogeneous lattices.
We first study a (conventional) Lieb lattice~\cite{Lieb1989}.
We take the lattice vectors as $\bm{a}_1^{\rm (L)} =(1,0)$, $\bm{a}_2^{\rm (L)} =(0,1)$,
and the coordinates of the sublattices are $\bm{r}_1^{\rm (L)} =(1/2,0)$, $\bm{r}_2^{\rm (L)} =(0,1/2)$, $\bm{r}_3^{\rm (L)} =(0,0)$. 
The lattice has site-dependent coordination numbers as
$z_1 = z_2 = 2$ and $z_3 = 4$. 
($z_\alpha$ is the coordination number of the sublattice $\alpha$.)
Notice that $x$ and $y$ are not sublattice-dependent, and are equal to one and zero, respectively. 

We explicitly show that we can tune the flat-band energy even on this lattice. 
To begin with, we consider the NN Hamiltonian given by
\begin{equation}
h^{\rm (L)}_1 (\bm{k}) =  t_1 \left(
\begin{array}{ccc}
0 & 0& h^{\mathrm{ (L,1)}}_{13}(\bm{k}) \\
0 & 0 & h^{\mathrm {(L,1)}}_{23}(\bm{k}) \\
h^{\rm (L,1)}_{13}(\bm{k}) &h^{\rm (L,1)}_{23}(\bm{k}) & 0\\
\end{array}
\right),
\end{equation}
with $h_{13}(\bm{k}) = 2 \cos \frac{k_x}{2}$ and $h_{23}(\bm{k}) = 2 \cos \frac{k_y}{2}$. 
The Hamiltonian has a flat eigenvalue $\lambda^{\rm (L,f)} = 0$ 
and the corresponding eigenfunction is 
\begin{eqnarray}
\psi^{\rm (L,f)} (\bm{k}) = \left( - \frac{\cos \frac{k_y}{2}}{\mathcal{N}^{(\mathrm{L})} (\bm{k})  }, \frac{\cos \frac{k_x}{2} }{\mathcal{N}^{(\mathrm{L})} (\bm{k}) } , 0 \right)^{\rm T},  \nonumber \\
\label{eq:Lieb_flat}
\end{eqnarray}
with $\mathcal{N}^{(\mathrm{L})} (\bm{k}) = \sqrt{ \cos^2 \frac{k_x}{2} + \cos^2 \frac{k_y}{2} }$.

The other two eigenvalues are given by
\begin{equation}
\lambda_1^{\rm (L,d)} (\bm{k}) = + 2 \sqrt{\cos^2\frac{k_x}{2} + \cos^2\frac{k_y}{2}}, 
\end{equation}
and
\begin{equation}
\lambda_2^{\rm (L,d)} (\bm{k}) = -2 \sqrt{\cos^2\frac{k_x}{2} + \cos^2\frac{k_y}{2}}, 
\end{equation}
thus they form a Dirac cone at M point, where they have a point contact with the flat band.  

Let us introduce the second-neighbor hopping, as 
\begin{equation}
h_2^{\rm (L)} (\bm{k}) = \left(
\begin{array}{ccc}
 h^{\rm (L,2)}_{11} (\bm{k})& h^{\rm (L,2)}_{12} (\bm{k})  & 0 \\
h^{\rm (L,2)}_{12} (\bm{k}) &h^{\rm (L,2)}_{22} (\bm{k}) & 0\\
0 &0 &h^{\rm (L,2)}_{33} (\bm{k})  \\
\end{array}
\right),
\end{equation}
with 
$h^{\rm (L,2)}_{11} (\bm{k})  = 2 \cos k_x $,
$h^{\rm (L,2)}_{12} (\bm{k}) = 2\left[\cos \left( \frac{k_x}{2} + \frac{k_y}{2}  \right) + \cos\left( \frac{k_x}{2} - \frac{k_y}{2} \right) \right] $,
$h^{\rm (L,2)}_{22} (\bm{k})  = 2 \cos k_y $,
and $h^{\rm (L,2)}_{33} (\bm{k})  = 2 (\cos k_x + \cos k_y )$. 
It is interesting to notice that, the block matrix for sublattice 1 and 2 is identical to 
the NN hopping matrix on a checkerboard lattice, which is a line graph.
This indicates that there exists a flat mode of $h_2^{\rm (L)} (\bm{k})$, which is, as we will see, identical to 
$\psi^{\rm (L,f)} (\bm{k}) $ in Eq. (\ref{eq:Lieb_flat}).

$h_2^{\rm (L)} (\bm{k})$ is not expressed by the quadratic form of $h_1^{\rm (L)} (\bm{k})$;
rather, it is expressed as 
\begin{equation}
h_2^{\rm (L)} (\bm{k}) = \left[h_1^{\rm (L)} (\bm{k}) \right]^2 
- \left(
\begin{array}{ccc}
2 && \\
&2 & \\
&&4 \\
\end{array}
\right),
\end{equation}
which reflects the fact that sublattice 3 has a larger coordination number than the other two sublattices. 
Nevertheless, the eigenvector of the flat mode of 
$h_1^{\rm (L)} (\bm{k})$, i.e., $\psi^{\rm (L,f)} (\bm{k})$, is also an eigenvector of $h_2^{\rm (L)} (\bm{k})$ with eigenvalue $-2$, 
since it does not have a weight on sublattice 3.  
Note that $\psi^{\rm (L,d)}_1(\bm{k})$ and $\psi^{\rm (L,d)}_2(\bm{k})$ are no longer eigenvectors after introducing the second-neighbor hopping. 

We show the band structures for several values of $t_1$ and $t_2$ in Fig. \ref{fig:Lieb}.
The analytical formula of the dispersion relations is presented in the appendix. 
As is in the case of the kagome and pyrochlore lattices, the band crossing does not occur for arbitrary $t_2$.
Indeed, for $|t_2| \leq \frac{|t_1|}{\sqrt{6}}$, the dispersive bands acquire the gap but they do not intersect the flat band:
Instead,  the lower dispersive band retains the touching point with the flat band at the M point  [Fig. \ref{fig:Lieb}(b)].
Meanwhile, for $|t_2| > \frac{|t_1|}{\sqrt{6}}$, the upper dispersive band intersects the flat band [Fig. \ref{fig:Lieb}(e)]. 

\subsection{Dice lattice}
\begin{figure}
\centering
\includegraphics[width=\linewidth]{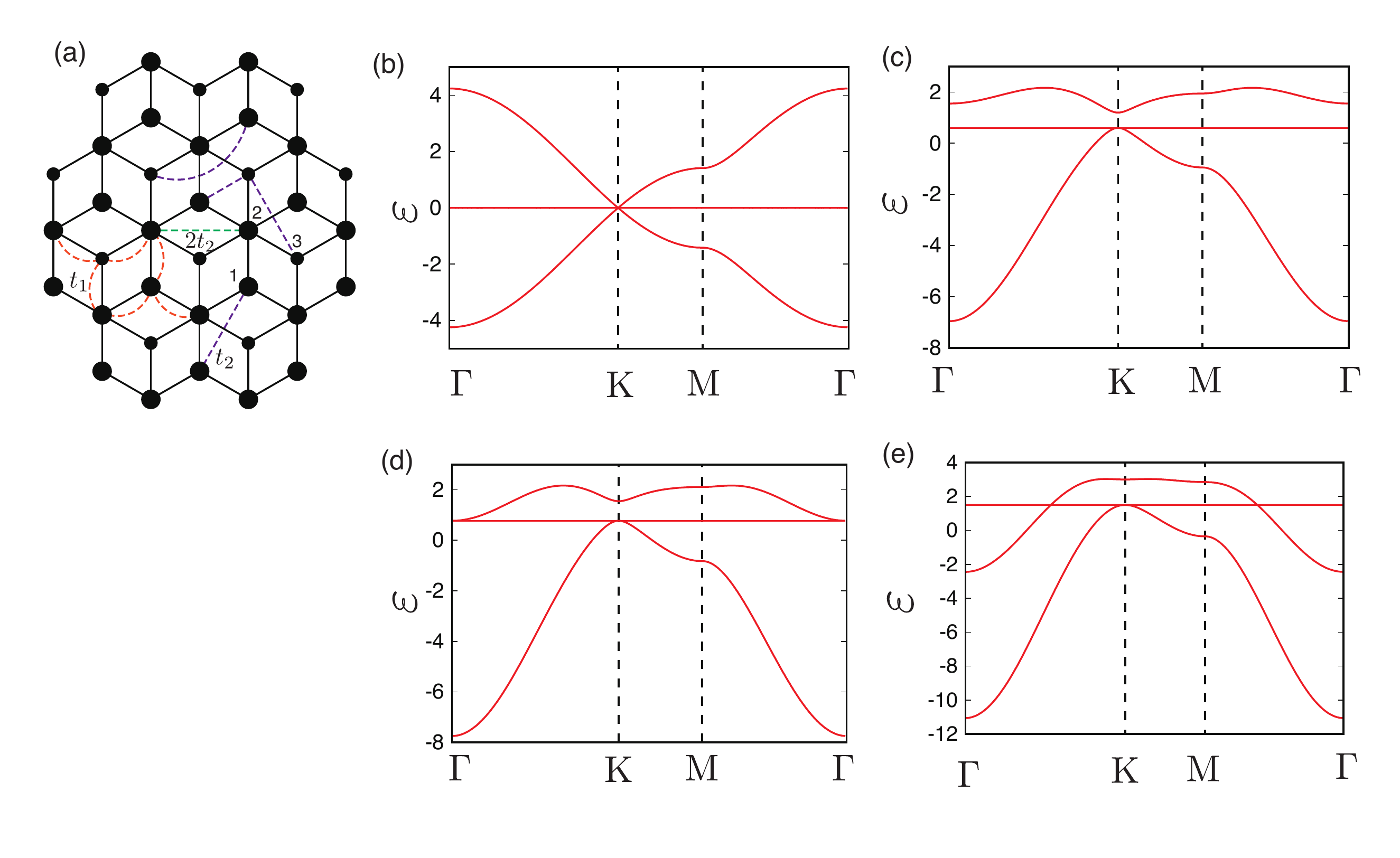}
\caption{
(a) A dice lattice.
Orange and purple dashed lines, respectively, denote the hopping processes with the hopping integrals $t_1$ and $t_2$.
For a green dashed line, the hopping integral is $2t_2$, as a consequence of the site-inhomogeneity (see main text). 
The sublattices are denoted by 1, 2 and 3.
The band structures for (b) $(t_1,t_2) = (-1,0) $, (c) $(t_1,t_2) = (-1,-0.2) $, (d) $(t_1,t_2) = (-1,-1/\sqrt{15})$, and (e) $(t_1,t_2) = (-1,-0.5)$.
The high-symmetry points in the Brillouin zone are given by 
$\Gamma = (0,0)$, K$=(\frac{4 \pi}{3},0)$, and M$=(\pi,\frac{\pi}{\sqrt{3}})$.
} 
\label{fig:Dice}
\end{figure}
We next study a dice lattice~\cite{Sutherland1986, Vidal1998, Vidal2001}, which has a trigonal symmetry. 
The lattice is constructed such that we add sites at the centers of hexagonal plaquettes on a honeycomb lattice;
each newly-added site has a finite hopping integral between only one of the two sublattices of an original honeycomb lattice, (say, 2).
Due to this choice of NN hopping, the coordination numbers differ from one sublattice to the others as $z_1  =z_3 = 3$, and $z_2=6$.
Furthermore, $x$ is also sublattice-dependent, as $x_{11} = x_{33} = x_{13}=1$, and $x_{22}=2$. 
($x_{\alpha \beta}$ is a number of the second-neighbor-hopping paths between sublattices $\alpha$ and $\beta$.)

We take the lattice vectors as $\bm{a}_1^{\rm (D)} = \left( \frac{1}{2} ,- \frac{\sqrt{3}}{2}\right)$,
$\bm{a}_2^{\rm (D)} = \left( \frac{1}{2} , \frac{\sqrt{3}}{2}\right)$, 
and the coordinates of the sublattices as
$\bm{r}_1^{\rm (D)} = \left( \frac{1}{2} ,- \frac{1}{2\sqrt{3}}\right)$,
$\bm{r}_2^{\rm (D)} = \left( \frac{1}{2} , \frac{1}{2\sqrt{3}}\right)$,
$\bm{r}_3^{\rm (D)} = \left(1,0\right)$. 
Then, the NN hopping matrix on this lattice is given as 
\begin{equation}
h_1^{\rm (D)} (\bm{k}) =\left(
\begin{array}{ccc}
0 &  h^{\rm (D,1)}_{12} (\bm{k}) & 0  \\
 h^{\ast \mathrm{(D,1)}}_{12} (\bm{k}) & 0 & h^{\rm (D,1)}_{12} (\bm{k})\\
0 & h^{ \ast \mathrm{(D,1)}}_{12} (\bm{k}) & 0 \\
\end{array}
\right),
\end{equation}
with $h^{\rm (D,1)}_{12} (\bm{k}) = e^{i \frac{k_y}{\sqrt{3}}} + 2 e^{-i\frac{k_y}{2\sqrt{3}}} \cos \frac{k_x}{2}$. 
$h_1^{\rm (D)} (\bm{k})$ has a flat eigenvalue $\lambda^{\rm (D,f)} = 0$, and 
the corresponding eigenfunction is
\begin{eqnarray}
\psi^{\rm (D,f)} (\bm{k}) = \left( \frac{h^{\rm (D,1)}_{12} (\bm{k})}{\sqrt{2}|h^{\rm (D,1)}_{12} (\bm{k})|},  0, - \frac{h^{\ast \mathrm{(D,1)}}_{12} (\bm{k})}{\sqrt{2}|h^{\rm (D,1)}_{12} (\bm{k})|}\right)^{\rm T},
\nonumber \\
\end{eqnarray}
which does not have a weight on sublattice 2. 
The other two bands are dispersive and given as,
\begin{equation}
\lambda_1^{\rm (D,d)} (\bm{k})  = 2 \sqrt{1+ 4 \cos^2  \frac{k_x}{2}  + 4 \cos \frac{k_x}{2}\cos \sqrt{3}k_y }
\end{equation}
\begin{equation}
\lambda_2 ^{\rm (D,d)} (\bm{k})  = - 2 \sqrt{1+ 4 \cos^2  \frac{k_x}{2}  + 4 \cos \frac{k_x}{2}\cos \sqrt{3}k_y }.
\end{equation} 
As is in the case of the Lieb lattice, they form a Dirac cone, where they touch the flat band. 

Now let us introduce the second-neighbor hoppings.
On this lattice, one needs a trick to obtain a desirable Hamiltonian, that is,
the second-neighbor hopping between neighboring sublattice 2 is twice as large as other second-neighbor hoppings.
This reflects the inhomogeneity of $x$. 
As a result, the second-neighbor hopping matrix we consider is given by 
\begin{equation}
h_2^{\rm (D)} (\bm{k}) = \left(
\begin{array}{ccc}
 h^{\rm (D,2)}_{11} (\bm{k})& 0 & h^{\rm (D,2)}_{13} (\bm{k})  \\
0 &2 h^{\rm (D,2)}_{11} (\bm{k}) & 0\\
h^{\ast \mathrm{(D,2)}}_{13} (\bm{k}) &0 &h^{\rm (D,2)}_{11} (\bm{k})  \\
\end{array}
\right),
\end{equation}
where $h^{\rm (D,2)}_{11} (\bm{k}) = 2 \left[ \cos k_x + \cos \frac{k_x + \sqrt{3}k_y}{2} + \cos \frac{k_x - \sqrt{3}k_y}{2} \right]$,
and $h^{\rm (D,2)}_{13} (\bm{k}) = 
e^{i \frac{\sqrt{3}k_x + k_y }{2\sqrt{3}}}
+e^{i \frac{- \sqrt{3}k_x + k_y }{2\sqrt{3}}}
+e^{-i \frac{ k_y }{ \sqrt{3}}}
+ e^{ i \frac{2 k_y}{ \sqrt{3}}}
+e^{ - i \frac{\sqrt{3}k_x +  k_y}{ \sqrt{3}}}
+e^{ i \frac{\sqrt{3}k_x -  k_y}{ \sqrt{3}}}$.
This satisfies the relation 
\begin{equation}
h_2^{\rm (D)} (\bm{k}) = [h_1^{\rm (D)} (\bm{k}) ]^2  -
\left(
\begin{array}{ccc}
3 & &\\
& 6& \\
&&3 \\
\end{array}
\right). 
\end{equation}
Again, since the eigenvector of the flat band does not have a weight at sublattice 2, 
it is also an eigenvector of $h_2^{\rm (D)} (\bm{k})$ with the eigenvalue $-3$. 

We plot the band structures for several values of $(t_1,t_2)$ in Fig. \ref{fig:Dice}.  
The analytical formula of the dispersion relations is presented in the appendix. 
As in the previous examples, there is a critical value of $|t_2|/|t_1|$ above which the band crossing between flat and dispersive band occurs. 
To be specific, the band crossing occurs for $|t_2|/|t_1| > \frac{1}{\sqrt{15}}$ [see Fig. \ref{fig:Dice}(e)]. 

\section{Application to an electric circuit \label{sec:circuit}}
\begin{figure}[b]
\centering
\includegraphics[width=0.95\linewidth]{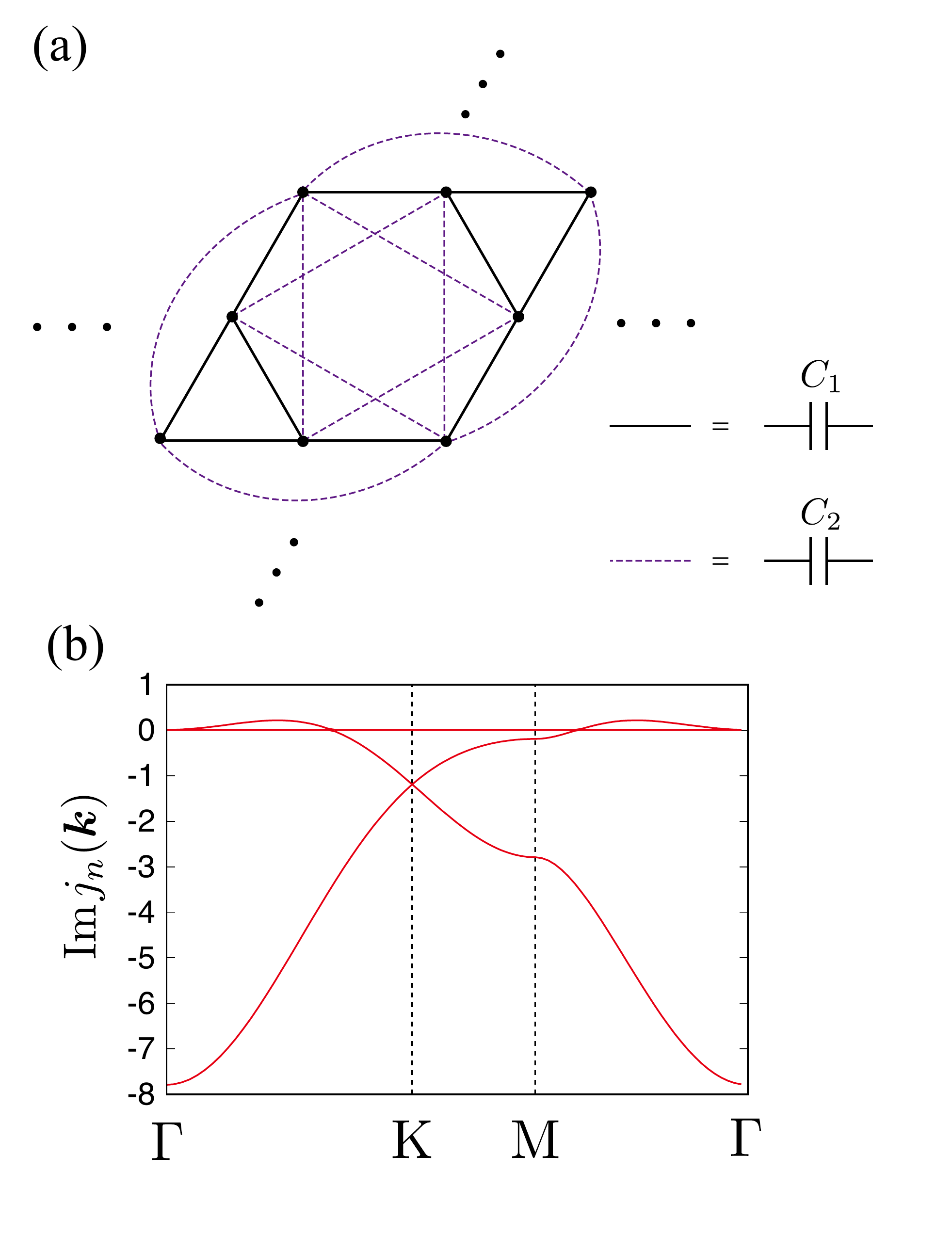}
\caption{
(a) Schematic figure for the electric circuit considered in Sec.~\ref{sec:circuit}.
Black filled circles denote the nodes of the circuit. 
Note that all nodes are connected to the ground by the inductors with inductance $L$.
Black solid lines and purple dashed lines denote the capacitors with capacitance $C_1$ and $C_2$, respectively. 
The dots denote the periodic boundary condition. 
(b) The band structure of the circuit Laplacian matrix 
for $(\omega C_1, \omega C_2,  \frac{1}{\omega L})= (1,0.3,7.79)$. 
} 
\label{fig:kagomecircuit}
\end{figure}
So far, we have shown that the Manhattan-distance-dependent hopping is crucial to tune the flat-band energies. 
In the solid-state systems, however, it is not always easy to control the hopping parameters. 
Nevertheless, recent developments of artificial materials,  
such as photonic crystals~\cite{Chern2015}, phononic crystals~\cite{Ni2017}, and electric circuits~\cite{Ezawa2018},
indicate that ``engineering" of the hopping parameters is possible in these systems, 
meaning that they will be an ideal platform to apply our method. 
In this section, we discuss one of those examples, namely an electric circuit with a kagome network. 

According to the modern theory of electric circuits~\cite{Cserti2011,Lee2018}, 
the relation between current ($I$) and voltage ($V$) in electric circuits
composed of periodic tiling of registers, capacitors, and inductors 
is described by using the circuit Laplacian matrix, 
which has a similar structure to tight-binding Hamiltonians in quantum mechanics. 
We emphasize that, since the hopping parameters of tight-binding Hamiltonians
exactly correspond to the resistance, capacitance, and inductance of the circuit elements in the circuit Laplacian formalism, 
the fine-tuning of the parameters can be achieved in these systems. 

To be concrete, consider the $LC$ circuit shown in Fig.~\ref{fig:kagomecircuit}. 
The nodes form a kagome lattice and they are connected each other by the capacitors. 
Note that all nodes are connected to the ground by the inductors with inductance $L$,
although they are omitted in the figure for simplicity. 
Suppose that the both the current and the voltage oscillate in time with an angular frequency $\omega$.
Then, the relation between $I_i (\omega)$ and $V_j (\omega)$ is written as 
\begin{equation}
I_i (\omega) = J_{i,j} (\omega)V_j (\omega).
\end{equation}
Here $J_{i,j} (\omega)$ is the circuit Laplacian matrix, and its explicit form
can be obtained by following Ref.~\onlinecite{Lee2018}:
\begin{eqnarray}
J_{i,j} (\omega) = \left( \frac{1}{i\omega L} + \sum_{\ell} i\omega C_{i,\ell} \right) \delta_{i,j} - i\omega C_{i,j},  
\end{eqnarray}
with $C_{i,j}$ being the capacitance between the node $i$ and $j$.
In the present setup, $C_{i,j} =C_1$ if $i$ and $j$ are connected by a NN bond,
$C_{i,j} =C_2$ if $i$ and $j$ are two-Manhattan-distance away, 
and otherwise $C_{i,j} =0$.

Furthermore, if the circuit satisfies the periodic boundary condition, 
we can perform Fourier transformation: 
\begin{equation}
\bm{I} (\omega, \bm{k} ) = J(\omega, \bm{k} ) \bm{V}(\omega, \bm{k} ),
\end{equation} 
where $\bm{I} (\omega, \bm{k} ) = \left[ I_1(\omega, \bm{k} ) , I_2(\omega, \bm{k} ) , I_3(\omega, \bm{k} )  \right]^{\mathrm{T}}$, 
$\bm{V} (\omega, \bm{k} ) = \left[ V_1(\omega, \bm{k} ) , V_2(\omega, \bm{k} ) , V_3(\omega, \bm{k} )  \right]^{\mathrm{T}}$, and 
\begin{eqnarray}
J(\omega, \bm{k} ) &=& 
\left[ 4i \omega (C_1 + 2 C_2 ) + \frac{1}{i \omega L }   \right] \hat{I}_3  \nonumber \\
&-&  i \omega C_1  h_1^{(\mathrm{K})} (\bm{k}) -  i \omega C_2 h_2^{(\mathrm{K})} (\bm{k}), \label{eq:circuit_lap}
\end{eqnarray}
where the explicit forms of $h_1^{(\mathrm{K})} (\bm{k})$ and $h_2^{(\mathrm{K})} (\bm{k})$ are 
shown in Sec.~\ref{sec:kagome}. 
Notice that $J(\omega, \bm{k} )$ is a non-Hermitian matrix, but $-i J(\omega, \bm{k} )$ is a Hermitian matrix,
thus the eigenvalues of $J(\omega, \bm{k} )$ are pure imaginary. 
The eigenvalues and eigenvectors of $J(\omega, \bm{k})$, $j_n(\omega, \bm{k})$ and $\psi_n(\omega,\bm{k})$, play a crucial role 
in determining the responses of the electric circuit. 
Namely, two-point impedance of the circuit between the site $(\bm{R}, \eta)$ and $(\bm{R}^\prime, \eta^\prime)$
can be written as~\cite{Lee2018}
\begin{eqnarray}
&Z^{\eta, \eta^\prime}(\bm{R}, \bm{R}^\prime )  \nonumber \\
= &  \sum_{\bm{k}, n}^{\prime}  \frac{| [\psi_n(\omega,\bm{k})]_{\eta}e^{i\bm{k} \cdot (\bm{R} + \bm{r}_{\eta})} - [\psi_n (\omega,\bm{k})]_{\eta^\prime}
e^{i\bm{k} \cdot (\bm{R}^\prime + \bm{r}_{\eta^\prime})}|^2}{j_n(\omega, \bm{k})}, \nonumber \\
\end{eqnarray}
where $\sum_{\bm{k}, n}^{ \prime}$ denotes summation over $\bm{k}$ and $n$ satisfying $|j_n(\omega, \bm{k})| \neq 0$. 
This indicates that the eigenmodes with $|j_n(\omega, \bm{k})| \sim 0$ play an important role in determining the two-point impedance.

In Fig.~\ref{fig:kagomecircuit}(b) we show the band structure, i.e., the momentum dependence of Im $j_n(\omega, \bm{k})$. 
Since $-i J(\omega, \bm{k} )$ has exactly the same structure as the tight-binding Hamiltonian of Eq. (\ref{eq:ham_general})
up the constant shift, our strategy to tune the flat band energy is applicable by tuning $C_2 /C_1$.
Indeed, we see that the flat band intersects the upper dispersive band. 
Furthermore, the flat-band energy can be tuned by inductance $L$ and the angular frequency $\omega$. 
In the present case, we fine-tune $ \omega L $ so that the flat band lies in the vicinity of zero.
Therefore, we expect an interesting $I$-$V$ response which arises from the mixture of the flat band and the dispersive band. 

\section{Conclusion \label{sec:conclusion}}
We have introduced a simple idea to tune the flat band energy by using farther-neighbor hoppings 
whose amplitudes are dependent on the Manhattan distance, instead of the real distance. 
Mathematically, this idea is based on the fact that, for a given matrix, polynomials of that matrix have the same eigenvectors as the original one
and its eigenvalues are given by the polynomial of the original ones. 
The merit of this method is that we do not need to fine tune many parameters to obtain the suitable band structure,
and that flat bands do not acquire a dispersion by the deformation of the Hamiltonian. 
We have also demonstrated that the proposed method is applicable to various lattices, 
including kagome and pyrochlore lattices, their breathing lattices, and a class of Lieb lattices. 

We expect that this method has broad potential applications to design suitable flat-band models.
As we have shown in Sec.~\ref{sec:circuit}, artificial materials are promising candidates, due to the tunability of the hopping parameters.
For solid-state systems, recent studies of first-principles calculations imply that carbon-based materials~\cite{Maruyama2016,Maruyama2017,Fujii2018,Fujii2018_2}, 
and pyrochlore oxides~\cite{Hase2018} will be promising candidates. 
Although the exact flatness will be spoiled by the additional hoppings in the real solid-state systems,
our method will serve as a good starting point 
to search the materials with nearly flat bands penetrating the dispersive bands,
which also show the intriguing physics. 
Studying the properties of those models, such as correlation effects, superconductivity, topological physics,
and effects of disorders~\cite{Vidal2001,Goda2006,Nishino2007,Chalker2010,Shukla2018,Bilitewski2018},
will be an interesting future problem. 

\acknowledgements
We thank Yasuhiro Hatsugai for fruitful discussions. 
T. M. would like to thank Yasumaru Fujii, Mina Maruyama, and Susumu Okada for enlightening discussions 
about Refs.~\onlinecite{Maruyama2016,Maruyama2017,Fujii2018,Fujii2018_2}.
This work was supported by Grants-in-Aid for Scientific Research, KAKENHI, JP17H06138 (T. M.),
 and JP15H05852 and JP16H04026 (M. U.), MEXT, Japan.
 
 \appendix
 \begin{widetext}
 \section{Analytical formulas for dispersion relations}
We summarize the dispersion relations for breathing kagome/pyrochlore lattices,
a Lieb lattice and a dice lattice in the presence of the second-neighbor hopping. 

\subsection*{Breathing kagome lattice}
In the NN hopping model, the eigenvalues and eigenvectors of dispersive bands can be obtained by using either 
a \lq \lq molecular orbital" method~\cite{Hatsugai2011} or a line-graph correspondence~\cite{Mizoguchi2018,Essafi2017}. 
Here we employ the latter, which is also applicable to the case with $t_2$.

We first introduce the incident matrix between the original kagome lattice 
and the dual honeycomb lattice: 
\begin{equation}
\tilde{T} (\bm{k}) = \left(
\begin{array}{ccc}
e^{i \varphi_1(\bm{k})} &  e^{i \varphi_2(\bm{k})} & e^{i \varphi_3(\bm{k})} \\
e^{-i \varphi_1(\bm{k})} &  e^{-i \varphi_2(\bm{k})} & e^{-i \varphi_3(\bm{k})} \\
\end{array}
\right), 
\end{equation}
where $\varphi_1(\bm{k})$-$\varphi_3(\bm{k})$ are defined in Sec. \ref{sec:kagome}.
Then, both the NN term and the second-neighbor term are expressed by $\tilde{T} (\bm{k})$~\cite{Mizoguchi2018,Essafi2017}:\begin{equation}
H = \sum_{\bm{k}}  (c^{\dagger}_{\bm{k},1} , c ^{\dagger}_{\bm{k},2},c^{\dagger}_{\bm{k},3} )
\left[\tilde{T}^{\dagger}(\bm{k}) 
\mathcal{D} (\bm{k})
\tilde{T} (\bm{k}) - (t_{1}^{\rm U} + t_{1}^{\rm D}  -  2 t_2)  \hat{I}_{3}
\right]
\left(
\begin{array}{c}
c_{\bm{k},1}\\
c_{\bm{k},2}\\
c_{\bm{k},3}\\
\end{array}
\right),
\end{equation}
where 
\begin{equation}
\mathcal{D} (\bm{k}) = \left(
\begin{array}{cc}
t_1^{\rm U}-2t_2 & t_2 F^{\rm (H)}(\bm{k}) \\
t_2  F^{\mathrm{(H)} \ast}(\bm{k})  & t_1^{\rm D} -2t_2\\
\end{array}
\right).
\end{equation}
$F^{\rm (H)}(\bm{k}) = e^{i \frac{k_y} {\sqrt{3}}} + 2 \cos \frac{k_x}{2} e^{-i\frac{k_y} {2 \sqrt{3}} }$ is 
the Fourier transformation of the NN hoppings on the dual honeycomb lattice. 

Now, an eigenvalue equation to solve is
\begin{equation}
\tilde{T}^{\dagger}(\bm{k}) 
\mathcal{D} (\bm{k})
\tilde{T} (\bm{k}) \psi (\bm{k})
= (\varepsilon + t_1^{\rm U} +t_1^{\rm D} -2t_2) (\bm{k})\psi(\bm{k}).  \label{eq:eigenK}
\end{equation}
To solve this, we define a two-component vector $\phi(\bm{k})$ such that
\begin{equation}
\phi(\bm{k}) = \tilde{T} (\bm{k}) \psi (\bm{k}). 
\end{equation}
Then, by multiplying $\tilde{T} (\bm{k})$ from left by Eq. (\ref{eq:eigenK}),
we obtain an eigenvalue equation for $\phi(\bm{k})$ as
\begin{equation}
\tilde{T} (\bm{k})  \tilde{T}^{\dagger}(\bm{k}) 
\mathcal{D} (\bm{k})
\phi (\bm{k})
=(\varepsilon + t_1^{\rm U} +t_1^{\rm D} -2t_2) \phi(\bm{k}).  \label{eq:eigenK2}
\end{equation}
Note that $\tilde{T} (\bm{k})  \tilde{T}^{\dagger}(\bm{k})$ is a $2\times 2$ matrix which is given as  
\begin{equation}
\tilde{T} (\bm{k})  \tilde{T}^{\dagger}(\bm{k}) = 
\left(
\begin{array}{cc}
3 & F^{\rm (H)}(\bm{k}) \\ 
F^{\mathrm{(H)} \ast}(\bm{k}) & 3\\
\end{array}
\right).  \label{eq:eigen_h}
\end{equation}
To obtain (\ref{eq:eigen_h}), we use the fact that $e^{2i \varphi_1(\bm{k})} + e^{2i \varphi_2 (\bm{k})} + e^{2i \varphi_3(\bm{k})} =F^{\rm (H)}(\bm{k})$. 

From(\ref{eq:eigen_h}), we see that the eigenvalue 
of the original problem, $\varepsilon^{\rm (BK)}(\bm{k})$, can be obtained by solving an eigenvalue equation of 
the $2\times 2$ matrix.  
By doing this, we finally obtain the dispersion relations as 
\begin{eqnarray}
 \varepsilon^{\rm (BK)}_{\pm} (\bm{k}) =   
\frac{ (t_1^{\rm U} + t_1^{\rm D} ) \pm \sqrt{9(t_1^{\rm U} - t_1^{\rm D})^2 + |F^{\rm (H)}(\bm{k})|^2 (t_1^{\rm U} + t_2)( t_1^{\rm D}  + t_2) }}{2} + ( |F^{\rm (H)}(\bm{k})|^2 - 4)t_2. \nonumber \\ 
\end{eqnarray}

\subsection*{Breathing pyrochlore lattice}
We can apply the same method to the breathing pyrochlore lattice with $t_2$.
Here we show the resulting eigenvalues of dispersive bands: 
\begin{eqnarray}
 \varepsilon^{\rm (BP)}_{\pm} (\bm{k}) =   
 (t_1^{\rm U} + t_1^{\rm D} ) \pm \sqrt{4(t_1^{\rm U} - t_1^{\rm D})^2 + |F^{\rm (D) }(\bm{k})|^2 (t_1^{\rm U} +2 t_2)( t_1^{\rm D}  + 2t_2) }+ ( |F^{\rm (D) }(\bm{k})|^2 - 6)t_2, \nonumber \\ 
\end{eqnarray}
with $F^{\rm (D) }(\bm{k}) = e^{-i \frac{k_x+k_y  +k_z}{8}} + e^{i \frac{-k_x+k_y  +k_z}{8}} +e^{i \frac{k_x-k_y  +k_z}{8}} +e^{i \frac{k_x+k_y  -k_z}{8}}$
being the Fourier transformation of the NN hoppings on the dual diamond lattice.

\subsection*{Lieb lattice}
In the following two cases, we obtain the eigenenergies 
by explicitly solving eigenvalue equations in two-dimensional space 
spanned by two dispersive modes of the NN Hamiltonian,
where we utilize that fact that the flat mode is unchanged when introducing the second-neighbor term. 

For the Lieb lattice, the dispersive bands have the following dispersion relations:
\begin{equation}
\varepsilon_{1}^{\rm (L)} (\bm{k}) = t_2 \left[ 4 \left( \cos^2  \frac{k_x}{2} +\cos^2  \frac{k_y}{2} \right)-3 \right] + \sqrt{ 4t_1^2 \left(\cos^2  \frac{k_x}{2} +\cos^2  \frac{k_y}{2} \right)  + t_2^2},
\end{equation}
and
\begin{equation}
\varepsilon_{1}^{\rm (L)} (\bm{k}) = t_2 \left[ 4 \left( \cos^2  \frac{k_x}{2} +\cos^2  \frac{k_y}{2} \right)-3 \right] -\sqrt{ 4t_1^2 \left(\cos^2  \frac{k_x}{2} +\cos^2  \frac{k_y}{2} \right)  + t_2^2}.
\end{equation} 
The upper dispersive band corresponds to $\varepsilon_{1}^{\rm (L)} (\bm{k})$. 
The critical value for $t_2$ at which the intersection between the flat and dispersive bands occurs is determined by the condition 
\begin{equation}
\varepsilon_{1}^{\rm (L)} (\bm{k}=0; t_1,t^c_2)  = - 2 t_2^c,
\end{equation}
which, as described in the main text, leads to $|t_2^c| = \frac{|t_1|}{\sqrt{6}}$. 

\subsection*{Dice lattice}
Next, we consider the dice lattice. 
The dispersive bands have the following dispersion relations:
\begin{equation}
\varepsilon_{1}^{\rm (D)} (\bm{k}) = t_2 \left( 2|h_{12}^{\rm (D,1)} (\bm{k})|^2 -\frac{9}{2} \right) + \sqrt {2 t_{1}^2 |h_{12}^{\rm (D,1)} (\bm{k})|^2 + \frac{9}{4}  t_2 ^2 },
\end{equation}
and
\begin{equation}
\varepsilon_{2}^{\rm (D)} (\bm{k}) =  t_2 \left( 2|h_{12}^{\rm (D,1)} (\bm{k})|^2 -\frac{9}{2} \right) -  \sqrt { 2 t_{1}^2 |h_{12}^{\rm (D,1)} (\bm{k})|^2 + \frac{9}{4}  t_2 ^2 },
\end{equation} 
where $h_{12}^{\rm (D,1)} (\bm{k})$ is given in the main text. 
The upper dispersive band corresponds to $\varepsilon_{1}^{\rm (D)} (\bm{k})$. 
The critical value for $t_2$ at which the intersection between the flat and dispersive bands occurs is determined by the condition 
\begin{equation}
\varepsilon_{1}^{\rm (D)} (\bm{k}=0; t_1,t^c_2)  = - 3 t_2^c,
\end{equation}
which, as described in the main text, leads to $|t_2^c| = \frac{|t_1|}{\sqrt{15}}$.

\end{widetext}

\end{document}